\documentclass[twocolumn,preprintnumbers,superscriptaddress,nofootinbib,aps,prd,floatfix]{revtex4}

\usepackage{graphicx}
\usepackage{amsmath,mathenv,bbm,amsfonts,color}
\usepackage{graphicx}
\usepackage{cases}
\usepackage{enumerate}

\newcommand{\Tr}{\text{Tr}}
\newcommand{\tr}{\text{Tr}}
\newcommand{\Br}{\text{Br}}

\newcommand{\gl}[1]{\eqref{#1}}
\newcommand{\CP}{$\mathcal{CP}$}
\newcommand{\zzf}{\Pi_{ZZ}(m_Z^2)}
\newcommand{\zzfn}{\Pi_{ZZ}(0)}
\newcommand{\azf}{\Pi_{AZ}(m_Z^2)}
\newcommand{\azfn}{\Pi_{AZ}(0)}
\newcommand{\aaf}{\Pi_{AA}(m_Z^2)}
\newcommand{\aafn}{\Pi_{AA}(0)}
\newcommand{\wwf}{\Pi_{WW}(m_W^2)}
\newcommand{\wwfn}{\Pi_{WW}(0)}
\newcommand{\alfa}{\alpha}

\newcommand{\SU}{\text{SU}}
\newcommand{\SO}{\text{SO}}
\newcommand{\U}{\text{U}}

\newcommand{\nn}{\nonumber}

\begin{document}

\title{Triplet Higgs boson collider phenomenology after the LHC}

\author{Christoph Englert} \email{christoph.englert@durham.ac.uk}
\affiliation{Institute for Particle Physics Phenomenology, Department
  of Physics,\\Durham University, Durham DH1 3LE, United Kingdom\\[0.3cm]}
\author{Emanuele Re} \email{emanuele.re@physics.ox.ac.uk}
\affiliation{Rudolf Peierls Centre for Theoretical Physics, Department
  of Physics,\\University of Oxford, Oxford, OX1 3NP, United Kingdom\\[0.3cm]}

\author{Michael Spannowsky} \email{michael.spannowsky@durham.ac.uk}
\affiliation{Institute for Particle Physics Phenomenology, Department
  of Physics,\\Durham University, Durham DH1 3LE, United Kingdom\\[0.3cm]}

\begin{abstract}
  ATLAS and CMS have discovered a Standard Model Higgs-like
  particle. One of the main discovery channels is the Higgs decay to
  two photons, which, at the moment, seems to be considerably enhanced
  over the Standard Model expectation. Models with additional charged
  matter coupling to the Higgs sector can enhance or decrease the
  $h\to \gamma \gamma$ branching ratio. We take this as motivation to
  confront the so-called Georgi-Machacek model of Higgs triplets with
  the results of recent searches for a SM Higgs boson performed at the
  LHC. We also investigate the model in regions of the allowed
  parameter space relevant for a SM-like phenomenology. The
  Georgi-Machacek model avoids tree-level issues of the $T$ parameter,
  while offering a vastly modified Higgs phenomenology compared to the
  Standard Model. This comes at the price of introducing another
  fine-tuning problem related to electroweak precision
  measurements. We investigate the collider phenomenology of the
  Georgi-Machacek model in the light of existing collider constraints
  beyond any effective field theory approximation and contextualize
  our findings with electroweak precision constraints.
\end{abstract}

\pacs{}

\preprint{DHCP/13/11}
\preprint{IPPP/13/22}
\preprint{OUTP-13-07P}

\maketitle


\section{Introduction}
ATLAS and CMS have reported on the discovery of a Standard Model
Higgs-like particle~\cite{orig} with a mass of approximately
126~GeV~\cite{:2012gk,:2012gu,newboundsa,newboundsc}. Bounds on production of SM Higgs-like
states with heavier masses have been established as low as
$\sigma\times \Br / [\sigma\times \Br]_{\text{SM}} \simeq
0.1$.

In the light of the late discovery which hints at deviations from the SM
expectation, attempts have been made to reconcile the excess in $h\to
\gamma\gamma$ in correlation with underproduction or ``spot-on''
production in the other decay and search channels. Given that there is
still consistency with the SM Higgs hypothesis within 1 to 2 sigma,
these results are easily misinterpreted. Preferring one model
over the other on basis of a better $\chi^2$ fit can be misleading:
Taking the tension of the Higgs mass of 126~GeV already within the SM
as an indicator of the SM's validity is certainly at odds with the
tremendous success that the SM has experienced so far, culminating in
the late LHC discovery.

Another way to check the validity of a certain model is to map the
uncertainty of the cross sections' extraction from data onto the
extended Higgs potential's parameters (and vice versa). Doing so, the
SM can in principle be recovered from the Higgs sector extension for
hypothetically accurate measurements without errors if the model can
approach a phenomenologically well-defined decoupling limit.

An enhancement of the $h\to \gamma\gamma$ rate can typically be
achieved by including additional charged states which are singlets
under QCD. This predominantly alters the decay branching ratios while
leaving the production cross sections unmodified apart from higher
order corrections and mixing effects. Higgs triplet extensions which
included color-neutral but up to doubly charged scalar particles are
therefore well-motivated model-building options to reconcile the
current observations~\cite{spencer,andrew,adam}. Constraints from
electroweak precision measurements and the correlation of the Higgs
candidate production cross section with exclusion bounds, that are
relevant for the remaining Higgs particles, are typically treated as a
nuisance in this context.

In simple triplet Higgs extensions, {\it i.e.} by introducing a
complex ${\bf{3}}_{1}$ under $\SU(2)_L\times \U(1)_Y$, the additional
Higgs bosons' phenomenology in SM-like search channels is typically
suppressed.  This comes from the fact that their couplings to the SM
fields are controlled by unitarity requirements being saturated by the
126~GeV candidate and by the small triplet's vacuum expectation value
(vev) needed for consistency with the $T$ parameter (the $W/Z$ mass
ratio)~\cite{Peskin:1990zt}. If the $T$ parameter issue is resolved at
tree-level by including another real triplet Higgs field under
$\SU(2)_L$, in what has become known as the Georgi-Machacek
model~\cite{Georgi:1985nv}, the Higgs phenomenology becomes more
involved, and the current measurements imply non-trivial constraints
on particle masses, couplings and the extended Higgs potential.

In the present paper we confront the Georgi-Machacek model of Higgs
triplets~\cite{Georgi:1985nv} with the measured results from LHC Higgs
searches of the 7 and 8 TeV runs. More precisely, we perform scans
over a simplified version of the model's potential to identify the
parameter region which is allowed in the light of current direct
searches and electroweak precision measurements.
Doing so, our approach is complementary to previous work by other
groups (see {\it e.g.}
Refs.~\cite{adam,spencer,Godfrey:2010qb,Cheung:2002gd,Chaichian:2009ts,Wang}). Instead
of using an effective lagrangian to extract information on the Higgs
couplings from data and then map these constraints on the model
parameter space, we use the full model and compare its predictions at
a given point in the parameter space with the observed data. We also
include constraints from electroweak precision measurements performed
during the LEP era, thus providing (to our knowledge) the most
detailed analysis of this model in the context of LEP and LHC collider
measurements hitherto.

\bigskip 

This work is organized as follows. We review the Georgi-Machacek (GM)
model in Sec.~\ref{sec:model} to make this paper self-contained and
comprehensive. In particular we introduce the potential that we scan
in the remainder of this work. In Sec.~\ref{sec:bounds} we discuss the
bounds which we take into account when scanning over the extended
Higgs sector phenomenology. We also give some technical details of our
implementations.  Sec.~\ref{sec:results} is finally devoted to
results, where we also detail the parameter choices of our scan. We
present our conclusions in Sec.~\ref{sec:conc}.

\section{The Model}
\label{sec:model}

It is well-known that Higgs triplets naively face compatibility issues
with electroweak precision data. This is due to the fact that a simple
triplet Higgs-extension of the Higgs potential leads to tree-level
custodial isospin violation, which is not present for complex
(symplectic) $\SU(2)_L$ doublets accidentally (as a consequence of
renormalizability and gauge invariance). This violation requires the
Higgs triplet's vev to be small compared to the weak scale in order to
obtain the experimentally observed $m_W/m_Z$ mass ratio.

Reconciling the $\rho=1+\alfa T$ parameter (at least at tree-level) in
a model with triplets requires more than a single triplet
field~\cite{Georgi:1985nv}. This can be seen by reminding ourselves of
how custodial isospin comes about for the SM doublet $\Phi$: If $\Phi$
transforms as a ${\bf{2}}$ under $\SU(2)_L$, then so does
$\Phi^c=i\sigma^{2}\Phi^\ast$. Consequently, the Higgs potential
depending only on $|\Phi|^2$ has a larger symmetry $\SU(2)_L\times
\SU(2)_R\simeq \SO(4)$, which breaks to $\SU(2)_D$ after the Higgs
obtains its vev. This ensures that the resulting electroweak gauge
boson masses are related by {\it{only}} the weak mixing angle. In
order to establish a $\SU(2)_R$ global symmetry in the Higgs potential
also in presence of $\SU(2)_L$ triplets, we need to enlarge the field
content with a real triplet such that $\SU(2)_R$ can act on the
complex triplet, its charge-conjugated version and the real triplet.

The GM model therefore introduces the Higgs fields
\begin{equation}
  \label{eq:higgsfields}
  \Phi=\left(\begin{matrix} \phi_2^{\ast} & \phi_1 \\ -\phi_1^{\ast} & \phi_2
    \end{matrix}\right), \quad 
  \Xi=\left(\begin{matrix} \chi_3^{\ast} & \xi_1 & \chi_1 \\
      -\chi_2^{\ast} & \xi_2 & \chi_2 \\
      \chi_1^{\ast} & -\xi_1^{\ast} & \chi_3 \\
    \end{matrix}\right)\,.
\end{equation}
In this notation, $\Phi$ is simply a SM-like Higgs doublet and $\Xi$
combines the complex $(\chi_1,\chi_2,\chi_3)$ and real
$(\xi_1,\xi_2,-\xi_1^*)$ triplets. Note that while symmetry breaking
with a correct $\rho$ parameter can be fully achieved with only $\Xi$,
the introduction of fermion mass terms still requires the presence of
a SM-like Higgs doublet.

The Higgs sector lagrangian that we consider in the remainder 
is
\begin{subequations}
  \label{eq:lag}
  \begin{multline}
    {\cal{L}}={1\over 2} \Tr \left[ D_{2,\mu} \Phi^\dagger D_2^\mu
      \Phi \right] + {1\over 2}\Tr \left[ {D}_{3,\mu} \Xi^\dagger
      {D}_3^\mu \Xi \right] - V(\Phi,\Xi) \\+ {\text{$\Phi$ Yukawa
        interactions}} \,,
  \end{multline}
  with the potential
  \begin{multline}
    \label{eq:pot}
    V(\Phi,\Xi)={\mu_2^2\over 2} \tr \left( \Phi^c \Phi \right) +
    {\mu_3^2\over 2} \tr \left( \Xi^c \Xi \right) + \lambda_1
    \left[\tr \left( \Phi^c \Phi \right)\right]^2 \\ + \lambda_2 \tr
    \left( \Phi^c \Phi \right) \tr \left( \Xi^c \Xi \right) +
    \lambda_3 \tr \left( \Xi^c
      \Xi\, \Xi^c \Xi \right)\\
    + \lambda_4 \left[\tr \left( \Xi^c \Xi \right)\right]^2 -
    \lambda_5 \tr \left(\Phi^c t_2^a \Phi t_2^b \right) \tr \left(\Xi^c t_3^a
      \Xi t_3^b \right) \,.
  \end{multline}
\end{subequations}
$D_2,D_3$ are the covariant derivatives for the doublet and triplet
representations, respectively, {\it{e.g.}}
\begin{equation}
  D_{2,\mu}\Phi=\partial_\mu\Phi + ig_w t^a_2 W^a_\mu \Phi -
  ig_y B_\mu  \Phi \, t^3_2 \,.
\end{equation}
Hypercharge $\U(1)_Y$ is embedded into $\SU(2)_R$ as in the SM. The
${\mathfrak{su}}(2)$ generators are $t^a_2=\sigma^a/2$ and
\begin{multline}
  t^1_3={1\over \sqrt{2}} \left( \begin{matrix}
      0 & 1 & 0 \\
      1 & 0 & 1 \\
      0 & 1 & 0 \end{matrix}\right)\,, \quad
  t^2_3={i\over \sqrt{2}} \left( \begin{matrix}
      0 & -1 & 0 \\
      1 & 0  & -1 \\
      0 & 1  & 0 \end{matrix}\right) \,,\\
  t^3_3=\left( \begin{matrix}
      1 & 0 & 0 \\
      0 & 0 & 0 \\
      0 & 0 & -1 \end{matrix}\right)\,.
\end{multline}

The potential in Eq.~\gl{eq:pot} is a simplified version of the
allowed terms documented in Refs.~\cite{Logan:2010en,Gunion:1989ci}:
the more general renormalizable and gauge invariant potential would
allow different vevs for the 2 triplet fields, and would also include
cubic terms (Eq.~(B3) of Ref.~\cite{Logan:2010en}). Terms of the
former type are $\SU(2)_R$-violating, and we avoid them by requiring
exact custodial invariance at tree level, as remarked in the
following. Terms with an odd number of scalars can be easily avoided
by means of a ${\mathbb{Z}}_2$ symmetry acting onto the triplet
fields.  Moreover our analysis is only indirectly affected by the
Higgs trilinear couplings (for measurement strategies of the latter
see Ref.~\cite{self}), and therefore our results are general enough to
assess the impact of Higgs measurements. Hence, our choice for
$V(\Phi,\Xi)$ should be thought of as a minimal ansatz, that captures
the important features of Higgs triplet phenomenology such as modified
Higgs branching ratios, production cross sections and Higgs mixing in
a well-defined way.

Switching off hypercharge gauging and the Yukawa interactions, the
lagrangian is manifestly invariant under $\SU(2)_L\times \SU(2)_R$.
Gauging a subgroup amounts to explicit breaking of custodial isospin,
and the effects of custodial isospin violation are steered by gauge
and Yukawa couplings. Hence a small $T$ parameter can be considered
natural~\cite{'tHooft:1979bh}.

Electroweak symmetry breaking is triggered by the scalar fields
developing vevs
\begin{equation}
  \label{eq:vevs}
  \left\langle\Phi\right\rangle=v_\Phi/\sqrt{2}\, 
  \mathbbm{1}\,, \quad \left\langle\Xi\right\rangle=v_\Xi \mathbbm{1}\,,
\end{equation}
as a consequence of minimizing the Higgs potential which also allows
us to eliminate $\mu_{2,3}^2$ and express them as functions of vevs
and $\lambda$'s. The vevs in Eq.~\gl{eq:vevs} are in accordance with
preserved custodial isospin. In principle we could have
$\left\langle\chi_3\right\rangle\neq \left\langle\xi_2\right\rangle$,
which would be induced in a $\SU(2)_R$-violating potential. In the
following we {\it{impose}} $\SU(2)_R$ invariance and such terms are
absent. This will have interesting consequences for the $T$ parameter.

The masses of the electroweak bosons $m_W,m_Z$ after symmetry breaking
follow the usual pattern of vev~$\times$~electroweak coupling, but the
electroweak scale is now generated by both the doublet and the triplet vevs
\begin{equation}
  \label{eq:vev}
(246~\hbox{GeV})^2\simeq v_{\text{SM}}^2= v_\Phi^2+8v_\Xi^2\,.
\end{equation}
As usual, it is useful to parametrize the relative contribution of
$v_\Phi$ and $v_\Xi$ to $v_{\text{SM}}$ via trigonometric functions:
\begin{equation}
  \label{eq:vevrot}
  \begin{split}
    \cos\theta_H =:&\;\,c_H={v_\Phi\over v_{\text{SM}}}\,, \\ 
    \sin\theta_H  =:&\;\,s_H={2\sqrt{2}v_\Xi\over v_{\text{SM}}}\,.
  \end{split}
\end{equation}

Compared to ``ordinary'' complex triplet Higgs extensions, as
considered recently in, {\it e.g.}, Ref.~\cite{Arhrib:2011vc,andrew}
to reconcile the $h\to \gamma\gamma$ enhancement, there is no
requirement to have a hierarchy $v_\Xi\ll v_\Phi$, {\it i.e.} the
phenomenology of electroweak bosons can be highly different from the
SM without being in immediate tree-level conflict with electroweak
precision data.  This is our main motivation to re-interpret the
current Higgs results in the context of the GM model.

\setcounter{footnote}{2}

The masses that arise from expanding the extended Higgs sector around
the minimum can be classified according to custodial isospin following
the above remarks. Eq.~\gl{eq:lag} yields a quintet, two triplets and
two singlets. The massless triplet contains the longitudinal degrees of
freedom of the electroweak gauge bosons, while the singlets mix as a
consequence of Eq.~\gl{eq:pot}. After diagonalizing the singlet mixing
of $H_\Phi,H_\Xi$\footnote{$H_\Phi\sim\Re\{\phi_2\}$ whereas $H_\Xi$
  is the linear combination of $\xi_2$ and $\Re\{\chi_3\}$ which
  yields a custodial singlet.}
\begin{equation}
  \begin{split}
    H_0 &= \phantom{-} c_q H_\Phi + s_q H_\Xi\,, \\
    H'_0 &= -s_q H_\Phi + c_q H_\Xi \,,
  \end{split}
\end{equation}
the Higgs mass spectrum reads
\begin{align}
  \begin{split}
    \label{eq:masses}
    m_{H_0}^2 &=  2 ( 2 \lambda_1 v_\Phi^2 + 2 ( \lambda_3 + 3 \lambda_4 ) v_\Xi^2 + m_{\Phi\Xi}^2 )\,, \\
    m_{H'_0}^2 &=  2 ( 2 \lambda_1 v_\Phi^2 + 2 ( \lambda_3 + 3 \lambda_4 ) v_\Xi^2 - m_{\Phi\Xi}^2 )\,, \\
    m_{H_3}^2&= {1\over 2} \lambda_5 (v_\Phi^2+8v_\Xi^2)\,,\\
    m_{H_5}^2&= {3\over 2} \lambda_5 v_\Phi^2 + 8 \lambda_3 v_\Xi^2\,,
  \end{split}
\end{align}
where
\begin{multline}
  m_{\Phi\Xi}^2 = \Big{[} 4 \lambda_1^2 v_\Phi^4 - 8 \lambda_1
  (\lambda_3 +3 \lambda_4) v_\Phi^2 v_\Xi^2 \\
   +  v_\Xi^2  
  \Big{(} 3 (2\lambda_2 - \lambda_5)^2 v_\Phi^2 
  + 4 (\lambda_3 + 3 \lambda_4)^2 v_\Xi ^2 \Big{)}
  \Big{]}^{1/2}\,,
\end{multline}
and
\begin{multline}
  \label{eq:sq}
  \sin \angle(H_\Phi,H_0) =:s_q\\=\frac{\sqrt{3}}
  {\sqrt{3+\Big{[}\frac{2\lambda_1 v_\Phi^2 -2(\lambda_3
        +3\lambda_4)v_\Xi^2 + m_{\Phi\Xi}^2}{{(2\lambda_2 - \lambda_5)}v_\Phi v_\Xi}\Big{]}^2}}\,.
\end{multline}
Note that the model contains a \CP~odd scalar $H_3^0$ which is the
neutral component of the massive custodial triplet (more precisely,
the \CP~odd scalar is the non-$Z_L$ combination of $\Im\{\phi_2\}$ and
$\Im\{\chi_3\}$).

Writing the full lagrangian in terms of the mass eigenstates we
recover the Feynman rules of the theory. For a general gauge ({\it
  i.e.}  including the Goldstone sector) we find ${\cal{O}}(500)$
potentially non-zero interaction vertices --- obviously too many to
discuss here. Most of the Feynman rules, however, arise from the Higgs
sector including the Higgs self couplings, Eq.~\gl{eq:pot}. We have
computed the Feynman rules
using~{\sc{FeynRules}}~\cite{Christensen:2008py} and validated the
output against an in-house Feynman rule extraction code. Some of the
Feynman rules can be found in earlier publications
~\cite{Gunion:1989ci,Logan:2010en,Cheung:2002gd,Godfrey:2010qb,spencer,Akeroyd:1995he}
and we also have performed analytical comparisons with these
results\footnote{In particular, to recover the Feynman rules for the
  Higgs self couplings listed in~\cite{Gunion:1989ci}, we notice that
  the $\lambda$'s couplings of~\cite{Gunion:1989ci} can be obtained
  from ours with the substitutions $\lambda_1\to\lambda_1+\lambda_3,
  \lambda_2\to 2\lambda_3+\lambda_4, \lambda_3\to 3\lambda_5,
  \lambda_4\to\lambda_2+\lambda_3-\lambda_5, \lambda_5\to
  2\lambda_4$. A number of Feynman rules in the GM
  phenomenology-pioneering work of Ref.~\cite{Gunion:1989ci} have been
  superseded in Ref.~\cite{spencer}.  }.

Before we discuss the bounds which we take into account for the
results of Sec.~\ref{sec:results} let us foreclose some
characteristic properties of the GM model
Eq.~\gl{eq:higgsfields}-\gl{eq:sq}:

\begin{enumerate}[1.)]
\item In contrast to models with a doublet and one complex triplet
  only, the GM model {\it a priori} allows for sizeable triplet
  vevs. Therefore, the couplings of the neutral Higgs mass eigenstates
  (in particular the two custodial singlets $H_0$ and $H'_0$) can
  highly differ from the SM Higgs couplings. We denote the universal
  ratios of the $hVV$ and $hf\bar{f}$ couplings of $H_0,H'_0$ with
  respect to the SM values by
  \begin{equation}
    \begin{split}
      \label{eq:ctcv}
      c_{t,H_0} &= \frac{c_q}{c_H}\,, \nn \\
      c_{v,H_0} &= c_q c_H + \sqrt{8/3} \,s_q s_H\,, \nn\\
      c_{t,H'_0}&= -\frac{s_q}{c_H}\,, \nn\\
      c_{v,H'_0}&= -s_q c_H + \sqrt{8/3} \,c_q s_H\,.
    \end{split}
  \end{equation}
  The above equations show that $c_v$ can be larger than unity, a
  typical property of models with weak isospin $j > 1/2$
  multiplets~\cite{Gunion:1984rv}. It is interesting to observe from
  the above expressions that values $|c_v|>1$ are possible only when
  the term with the prefactor $\sqrt{8/3}\sim 1.6$ dominates over the
  first term. Values of $|c_v|$ very close to one can be obtained
  either when the first term is very close to unity (which means very
  small triplet vev and minimal mixing in the singlet sector, {\it
    i.e.}~$c_q$ very close to one or zero) or when there is a
  compensation between the two terms in $|c_v|$. This compensation can
  potentially take place only when the second term is not too
  suppressed, {\it i.e.} an enhanced $|c_v|$ does not occur for $0\leq
  s_H \ll 1$.

\item Another very interesting feature of the GM model shows up for
  vanishing mixing of the singlet states 
  In this situation the mass eigenstates are by definition aligned
  with the $H_\Phi, H_\Xi$ states. If we take the limit $s_H (c_H)
  \simeq 0$, which corresponds to breaking electroweak symmetry mainly
  with the doublet (triplet) vev, we get a massless scalar in the
  spectrum. This state always corresponds to $H'_0$ and, as we will
  see, it has important consequences for the model's phenomenology.

  Away from the edges of the parameter space ({\it i.e.}~when neither
  $s_H$ nor $c_H$ are very small, and $\lambda$'s are all of the same
  order), the above mechanism can be rephrased by saying that the
  lighter custodial singlet Higgs boson is usually the one which is
  less related to electroweak symmetry breaking~\cite{Gunion:1989ci}.
  When the mixing in the singlets sector is non-negligible, the mass
  spectrum becomes effectively less affected by this property, and the
  mass of the lightest Higgs boson can be made sizeable: the allowed
  size of this misalignment is encoded in our results.

  Moreover, despite of these considerations, all states clearly
  contribute to unitarization of longitudinal gauge boson scattering
  (the longitudinal gauge bosons are mixtures of the triplet $\Xi$ and
  doublet $\Phi$), and this has interesting implications for the
  electroweak precision constraints, especially in the scenario
  discussed in Sec.~\ref{sec:noprime}.

  For a more general potential that also violates $\SU(2)_R$
  invariance, the $s_H\to 0$ limit does not necessarily involve
  $m_{H'_0}\to 0$, {\it cf.} \cite{Logan:2010en}. For these more
  general potentials there always exist parameter choices that involve
  light scalars in the spectrum. The correlation with $s_H$, however,
  is not so obvious. Generically the light scalars follow from
  parameters with a small $\SU(2)_R$ breaking. Since $\rho\simeq 1$ is
  well-established experimentally, invariance under global $\SU(2)_R$
  is well-established too, and the allowed isospin violation can be
  treated as a perturbation which should restore, to large extent, the
  correlation between vanishing $s_H$ and $m_{H'_0}\to 0$, for
  phenomenologically allowed choices. In fact, the limit $v_{\Xi}\to
  0$ in Eq.~\gl{eq:pot} is tantamount to $\mu_3^2\to
  (-2\lambda_2+\lambda_5)v_{\text{SM}}^2$, which means that the
  custodial singlet part of the Higgs potential effectively ``sees'' a
  vanishing mass term, while this is not the case for the custodial
  triplet and quintet.
\end{enumerate}
\section{Bounds}
\label{sec:bounds}

The implications of the GM model for the Peskin-Takeuchi
$S,T,U$ parameters~\cite{Peskin:1990zt},
\begin{widetext}
  \begin{align}
    \label{eq:stu}
    \begin{split}
      S&= {4 s_w^2c_w^2 \over \alfa} \left(\frac{\zzf - \zzfn}{m_Z^2}
        - {c_w^2-s_w^2\over c_ws_w}{\azf-\azfn\over m_Z^2} -
        {\aaf-\aafn\over m_Z^2}\right) \,,\\
      T&= {1\over \alfa}\left( {\wwfn\over m_W^2} - {\zzfn\over m_Z^2}
        - {2 s_w\over c_w} {\azfn\over m_z^2}-{s_w^2\over c_w^2}{\aafn\over m_Z^2}\right)\,, \\
      U&= {4 s_w^2\over \alfa} \left( {\wwf-\wwfn\over m_W^2} - c_w^2
        {\zzf - \zzfn\over m_Z^2 } \right.
      \notag\\
      &\hspace{5cm}- \left. s_w^2{\aaf-\aafn\over m_Z^2} -2 s_w c_w
        {\azf-\azfn\over m_Z^2} \right) \,,
    \end{split}
  \end{align}
\end{widetext}
have been pioneered in Ref.~\cite{Gunion:1990dt}. 

The parameters $\Pi_{XY}$ denote the gauge boson polarization
functions which are obtained from the two-point functions~$X\to
Y$~($X,Y=A,Z,W$),
\begin{equation}
  \label{eq:twopoint}
  \Pi_{XY}(Q^2)\, \sim \parbox{3.5cm}{\includegraphics[width=3.5cm]{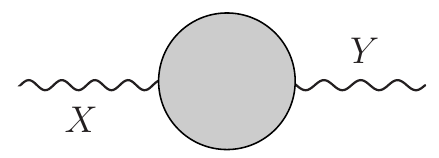}}\,,
\end{equation}
$\alpha$ is the fine structure constant and $s_w, c_w$ are the sine
and cosine of the weak mixing angle as usual. The blob denotes all
particle dynamics which enters the polarization functions in the
corresponding model.

At this stage, it is important to note a peculiarity of the $T$
parameter that arises in the tree-level $\SU(2)_R$ symmetric GM
model. Although the identification between the real and complex
triplets' vevs yields the correct tree-level $W/Z$ mass ratio, the
situation becomes more involved at the one-loop. The real triplet's
vev requires the introduction of a counter term (related to the weak
mixing angle renormalization) that could be naively unexpected from
the $\SU(2)_R$-symmetric tree-level
lagrangian~\cite{Gunion:1990dt}. This counter-term mends a residual UV
singularity in the $T$ parameter. In this sense, $\SU(2)_R$ invariance
introduces an additional naturalness problem in addition to the
conventional hierarchy problem, namely why the $T$ parameter remains
small. This finding is somewhat similar to the original $T$ parameter
problem in $\SU(2)_R$-violating triplet extensions.

Performing the full one-loop calculation with the Feynman rules
described in the previous section, we find that, after introducing a
corresponding $T$ parameter counter term, our calculation is
manifestly free of UV divergencies; no residual UV divergencies appear
in neither the $S$ nor in the $U$ parameter, in accordance with the
results of Ref.~\cite{Gunion:1990dt}.

As pointed out in Ref.~\cite{Gunion:1990dt} the GM model picks up a
quadratic $T$ parameter divergence which introduces a fine-tuning
problem analogous to the Higgs mass. The $T$ parameter constraint is
very stringent and will probe the finite logarithms. If these
logarithms are of order one, a large one-loop $T$ parameter can
decrease again when the full perturbative expansion is
known\footnote{See {\it e.g.}  Ref.~\cite{wetterich} for a solution to
  the Higgs naturalness problem in this context}. We therefore trace
the effect of imposing $S$ independent from $T$ in
Sec.~\ref{sec:results}. When imposing only the $S$ parameter
constraint we set $T=0$. This can in principle be interpreted as a
situation where a large $T$ shows up as relic of perturbation theory.

At this point it is worthwhile to comment on the applicability of
$S,T,U$ for the present model. Estimating new physics effects by means
of $S,T,U$ assumes, among other approximations~\cite{Barbieri:2004qk},
that the new physics scale $\Lambda$ is larger than $m_Z$. Identifying
$m_{H_0}$ as the 126 GeV candidate we can face spectra for which
$m_{H'_0}$ is close to or even below the $Z$ threshold.  In this case,
the expansion of the vacuum polarization functions in terms of
$Q^2/\Lambda^2$ on which $S,T,U$ rely\footnote{We would like to remind
  the reader that small allowed $S,T,U$ values are necessary but not
  sufficient conditions for consistency with electroweak precision
  data. The latter requires carrying out an analysis including
  model-dependent radiative corrections to at least next-to leading
  order.} could be ill-defined if the coupling $|c_{v,H'_0}|$ is
large. Then, the light scalar cannot be considered a small
perturbation compared to the heavier states' impact. Including the LEP
constraints on light SM-like Higgs production ($m_H\leq
114~\text{GeV}$), which probe combined mixings down to $c^2_v \sim
0.01$, the $S,T,U$ bounds are superseded by the direct exclusion from
$e^+e^-\to hZ$ searches. If the mixing is too large, the direct LEP
constraints remove the parameter point from the list of valid points,
especially when the resulting $\Delta S$ falls into the allowed
region, and our theoretical predicament is resolved
phenomenologically.

Additional constraints that in principle reduce the model's allowed
parameter space follow from non-oblique corrections, {\it e.g.}
corrections the $Zb\bar b$ vertex, and flavor physics. Modifications
of the flavor sector are induced by ``rotating in'' the custodial
triplet $(H_3^+,H^0_3,H_3^-)$ via Goldstone mixings in the extended
space of Higgs fields, Eq.~\gl{eq:higgsfields}, and by the singlets'
mixing.

These mixings are parametrically controlled by $c_q, s_q, c_H$ and
$s_H$. The qualitative modifications in the flavor sector in
comparison to the SM are similar to a two-Higgs doublet model without
flavor changing neutral currents~\cite{2hdm}. We expect that
consistency with flavor constraints can be achieved without too large
quantitative changes induced (see {\it e.g.}
Ref.~\cite{adam}). Especially for scenarios with tuned $T$, which, as
we will see in Sec.~\ref{sec:results}, are consistent for small $s_H$,
we can expect flavor physics to be SM-like. Similar arguments hold for
the non-oblique corrections, {\it e.g.} the $Zb\bar b$ vertex does not
receive corrections from the quintet states at the one-loop
level~\cite{Haber:1999zh}.

Beyond indirect constraints such as electroweak precision
measurements, there are already direct constraints from searches
$H_5^{\pm\pm} \to \ell^\pm\ell^\pm$ ($\ell =e,\mu$) by both ATLAS and
CMS~\cite{searchhpp}. Such a decay can be prompt by the introduction
of a Majorana-type interaction term of the triplet and the leptons and
constrains a different sector of the model. The bounds of
Ref.~\cite{searchhpp} assume coupling strengths of order 1 to derive
lower mass limits on the quintet mass. While this assumption is
reasonable to obtain a well-defined limit in hypothesis tests,
especially for a heavy quintet $m_{H_5}\gtrsim 2 m_W$ this is not
reasonable. For large $m_{H_5}$, we need to include a partial decay
width which is dominated by longitudinally polarized $W$ bosons. This
leads to a tree-level dependence of the partial decay width
$\Gamma(H_5^{\pm\pm}\to W^\pm W^\pm) \sim m_{H_5}^3/m_W^2$. Just like
in the SM, the decay to light fermions scales $\Gamma(H_5^{\pm\pm}\to
\ell^\pm \ell^\pm) \sim m_{H_5}$, and the decay to same-sign leptons
quickly becomes subdominant. Searches for same-sign inclusive
dileptons exist~\cite{Aad:2011vj} but are typically designed to cope
with non-resonant squark-gluino
production~\cite{Aad:2011vj,GoncalvesNetto:2012yt}. Additional
constraints from leptonic final states can be obtained from
measurements of lepton flavor violating $\mu^\pm\to
e^+e^-e^\pm$~\cite{lepflv,Akeroyd:2009nu,Konetschny:1977bn}.

We do not include these constraints in detail in the following. We
however note that dedicated limits can be obtained from an analysis of
the $\ell^\pm\ell^\pm jj$ final state for $H_5^{\pm\pm}$ production
via Weak Boson Fusion (WBF).


\section{Results}
\label{sec:results}
We now turn to the results of our scans over the Higgs potential. The
role of the Higgs candidate can in principle be played by all
uncharged Higgs states $H_0,H'_0,H_3^0,H_5^0$. In the light of recent
results, however, it is unreasonable to study $H_3^0$ and $H_5^0$ in
further detail:
\begin{itemize}
\item $H_3^0$ is a \CP~odd state which does not couple to the massive
  vector bosons. The measurement of the Higgs candidate in the $ZZ$
  channel~\cite{:2012gk,:2012gu} which also disfavors pure \CP~odd
  quantum numbers from angular distributions in the leptonic decay
  channels~\cite{CMSCP} removes $H_3^0$ from our list of
  candidates. Note that constraints on \CP~odd scalars directly
  constrain $\lambda_5$, Eq.~\gl{eq:masses}. Furthermore, a 126 GeV
  $H_3^0$ is also disfavored from flavor physics and non-oblique
  corrections~\cite{Haber:1999zh}.
\item Even though $H_5^0$ has the right quantum numbers and vertex
  structures to mimic the branchings to the $\gamma\gamma,\ ZZ$ and
  $WW$ final states, the quintet is fermiophobic as a consequence of
  the $\SU(2)_R$ symmetry. This removes gluon fusion~\cite{gf} as
  production mode and leaves inclusive weak boson fusion~\cite{wbf} as
  the largest production cross section for $H_5^0$. Weak boson fusion
  for inclusive cuts is typically suppressed by one order of magnitude
  compared to gluon fusion~\cite{Dittmaier:2011ti}. Reconciling the
  measurements in $ZZ,WW,\gamma\gamma,\tau^+\tau^-$ with the observed
  rates is therefore impossible, and $H_5^0$ cannot play the role of
  the 126 GeV candidate, when considering recent limits on the total
  Higgs width~\cite{atlconf}.  
\end{itemize}

From Eq.~\gl{eq:masses} it is manifest that $H'_0$ is always lighter
than $H_0$, and hence $H'_0$ is the natural candidate to play the role
of the observed Higgs. As already mentioned in Sec.~\ref{sec:bounds},
we will also study the scenario where there is a lighter uncharged
scalar state in the mass spectrum, which we can achieve by fixing
$m_{H_0}=126$ GeV. However we will see that this scenario is 
disfavored.

Although partial motivation of this work is to explore the GM model's
potential to enhance the $h\to\gamma\gamma$ branching ratio via the
presence of extra matter, it is clear from the Higgs couplings not
fixed to the SM values (Eq.~\gl{eq:ctcv}) that the $\gamma\gamma$
branching ratio in this model is also affected by these modified
tree-level couplings. This means that it is indeed possible to find
regions in the parameter space where $\Br(H\to\gamma\gamma)$ is very
close to the SM also without forcing the charged higgses to be very
heavy. We will briefly explore this possibility too, by looking for
points that reproduce the Higgs signal and at the same time have a
$\gamma\gamma$ branching ratio close to unity. Precise fits on Higgs
couplings will eventually tell whether this is a feasible scenario,
although recent studies by several collaborations favor regions with
$|c_t|,|c_v|\sim 1$.

The cross section limits for the uncharged Higgs fields are adopted
from the most recent LHC measurements~\cite{newboundsa,newboundsc} and
from the combined LEP constraints~\cite{LEPred}. In particular for
Higgs masses smaller then the LEP direct bound, we use the LEP bound
on the $HZZ$ coupling.

Signal strengths are defined as follows:
\begin{widetext}
  \begin{equation}
    \label{eq:muvalue}
    \mu_{h\to XX}=\frac{\sigma(pp\to h)\times \Br(h\to XX)}
    {\sigma_{\text{SM}}(pp\to h)\times \Br_{\text{SM}}(h\to XX)}\,,
  \end{equation}
  and for this paper we define the combined signal strength as
  \begin{equation}
    \label{eq:comb}
    \mu_{h}=\frac{\sigma(pp\to h)\times [\Br(h\to WW)+\Br(h\to ZZ)
      +\Br(h\to \gamma\gamma)]}{\sigma_{\text{SM}}(pp\to h)\times 
      [\Br_{\text{SM}}(h\to WW)+\Br_{\text{SM}}(h\to ZZ)+\Br_{\text{SM}}(h\to \gamma\gamma)]}
  \end{equation}
\end{widetext}
since ATLAS and CMS use the categories $WW$, $ZZ$, $\gamma\gamma$,
$\tau\tau$ and $Vb\bar{b}$ to obtain the exclusion bounds from data,
and the latter two categories give a marginal contribution to $\mu_h$,
because of non-observation of the Higgs and large experimental errors in those
channels. Eq.~\gl{eq:comb} does not take into account the different
channels' sensitivity which is beyond the scope of this work.

We scan $\lambda_i\in [-4\pi,4\pi],~i=1,\dots,5$ and $s_H~\in~[0,1]$,
which also implies that since $\lambda_5$ is bounded from above, in
our scan the maximum value allowed for $m_{H_3}$ is $\sim$ 600 GeV. To
generate parameter points with $m_{H'_0} (m_{H_0})=~126\pm 1$~GeV more
efficiently, we first generate the mass of the $126$~GeV state, and
then compute $\lambda_1$ ($\lambda_1$ and $\lambda_2$) using
Eq.~\gl{eq:masses}.  The generated parameter points are, hence, not
flat and should by no means be understood as probability distributions
of parameter points that pass the requirements.

\begin{figure*}[!t]
  \includegraphics[width=0.45\textwidth]{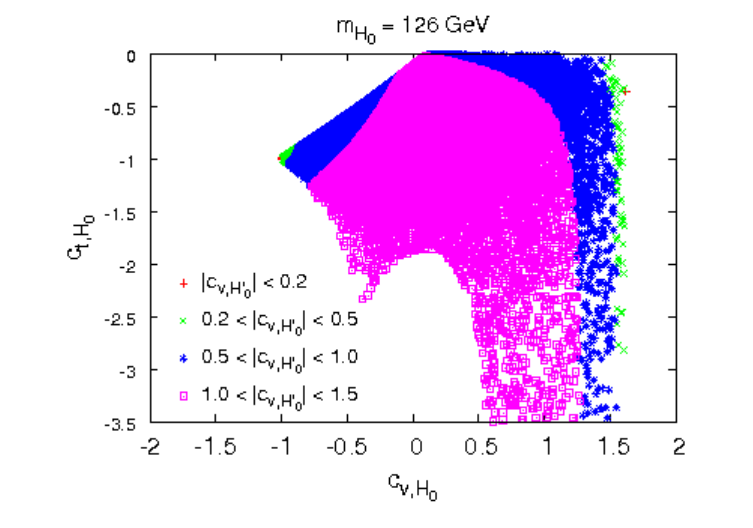}
  \hspace{1cm}
  \includegraphics[width=0.45\textwidth]{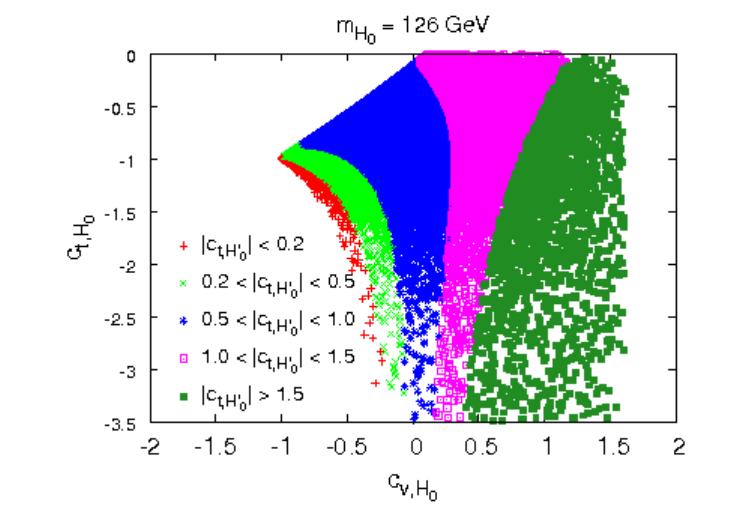}
  \caption{\label{fig:h0cvct} $(c_{v,H_0},c_{t,H_0})$ correlation for
    $m_{H_0}\simeq 126$~GeV. We do not impose any other additional
    constraint.}
\end{figure*}

All the LHC measurements of the Higgs-candidate properties, in
particular from the $WW$, $ZZ$ and $\gamma\gamma$ categories, point
towards an Higgs phenomenology with $|c_t|$ and $|c_v|$ not
dramatically different from the SM
values~\cite{rec,plehnrauchlhc,adam}. Moreover, the combined signal
strengths from CMS and ATLAS are $\mu_{h,{\text{CMS}}}\simeq 0.9 \pm
0.2$~\cite{newboundsc} and $\mu_{h,{\text{ATLAS}}} \simeq 1.3 \pm
0.3$~\cite{newboundsa}.  We will take the ATLAS measurement as the
paradigm of an enhanced $h\to \gamma\gamma$ production rate, but will
also comment on the model's capability to reproduce the consistency
with the SM as observed by CMS.

Since the theoretical expectation for these values is dominated by
$\mu_{h\to WW}\sim |c_t c_v|^2$, it is reasonable to restrict our
scans to the regions where both $|c_t|$ and $|c_v|$ are larger than
0.8.\footnote{The $\gamma \gamma$ excess, although phenomenologically
  very important if confirmed, cannot be responsible for a 20\%
  enhancement of the global $\mu_h$ value, being a potentially large
  effect in a rare branching ratio.}. Unless otherwise stated, all
results have been obtained with the aforementioned condition
explicitly imposed.

We use the exclusion contours by ATLAS\footnote{While the exclusion
  limits of both ATLAS and CMS do not coincide at face value, they
  quantitatively follow the same pattern.}\cite{:2012gk}. In order
to reproduce the observed LHC discovery signal we need an excess of
$\sim 25 \%$ for the total signal strength compared to the SM
hypothesis. We therefore typically need either $|c_t|$ or $|c_v|
>1$. The cross section in the discovery channels scales as $c_t^2$ and
the $VV$ branching ratios as $\sim c_v^2/c_t^2$. Therefore $c_t$ is
typically less constrained in our scan, also because no statistically
significant observation in the fermionic channels has been made so
far. The approximate scaling for the $VV$ branching ratio mentioned
above follows from the observed ``SM-likeness'' of the 126 GeV Higgs
boson, which in turn implies the decay to bottom quarks to be the
dominant contribution to the total decay width. Notice however that in
our approach we have taken into account all the possible changes to
the Higgses' total widths due to non SM-like couplings and to the
extended Higgs sector. In the scenario with an uncharged scalar
lighter than the 126 GeV state, a too large deviation of the total
width when $H_0\to H'_0 H'_0$ opens up is limited by the experimental
observation of the individual signal strengths being in good agreement
with the SM values.

To estimate the signal strengths for $H_0$, $H'_0$ and $H_3^0$ and
compare with the exclusion bounds, we need their production cross
sections and branching ratios. The branching ratios and total widths
have been obtained interfacing~{\sc{FeynRules}} with a modified
version {\sc{Bridge}}~\cite{Meade:2007js}, where we included the loop
functions needed to evaluate the amplitudes $h\to \gamma \gamma$ via
scalar, fermionic and vectorial loops. We have also included the more
important QCD corrections to the $h\to b\bar{b}$, $h\to c\bar{c}$ and
$h\to g g$ decays~\cite{Djouadi:2005gi}, and cross-checked the
implementation against the SM partial widths quoted
in~\cite{Dittmaier:2011ti}. {\sc{Bridge}} also computes the branching
ratios for off-shell decays.

The leading-order inclusive cross sections $\sigma(gg\to h)$ have been
computed using an adapted version of the {\sc{Powheg-Box}}
program~\cite{Alioli:2010xd}. Using LO cross sections is a good
approximation since higher order QCD corrections play a minor role as
far as signal strengths are concerned. In doing so, we have also
neglected the contribution from Higgs production via WBF, which plays
a subleading role in our study. We leave a detailed discussion of the
GM's WBF phenomenology to the future.

To impose our electroweak precision criteria we use the $S,T,(U\equiv
0)$ fits at 95\% confidence level of
Ref.~\cite{lepepw}\footnote{During the course of this work, these fits
  have been updated including the Higgs
  measurements~\cite{Baak:2012kk}. The differences compared to
  Ref.~\cite{lepepw} are quantitatively small and not relevant for our
  results.}. We classify a point in the parameter space as ``good''
with respect to the $m_{H_0},m_{H'_0} \simeq 126$~GeV triplet
hypotheses according to three main different requirements as already
alluded to before.

\begin{enumerate}[(i)]
\item {\it{loose electroweak precision:}} $S$ within the 95\% CL
  ellipse, no constraint on $T$.  We understand this case as $T\equiv
  0$, along the lines of Sec.~\ref{sec:bounds}.
\item {\it{enforced electroweak precision:}} Both $S$, $T$ strictly in
  the 95\% CL ellipse.
\item {\it{bounds from global direct searches and tagged categories:}}
  in this case we reproduce the $\mu_h$ value for the Higgs signal
  within the quoted error of $\simeq$ 25\% and we do not violate the
  exclusion bounds for the other neutral Higgses. We also reproduce
  $\mu_{h\to WW}$ and $\mu_{h\to\gamma\gamma}$ within the $1 \sigma$
  error band.
\end{enumerate}

\begin{figure*}[p!]
  \centering
  \hspace{0.2cm}~
  \includegraphics[width=0.4\textwidth]{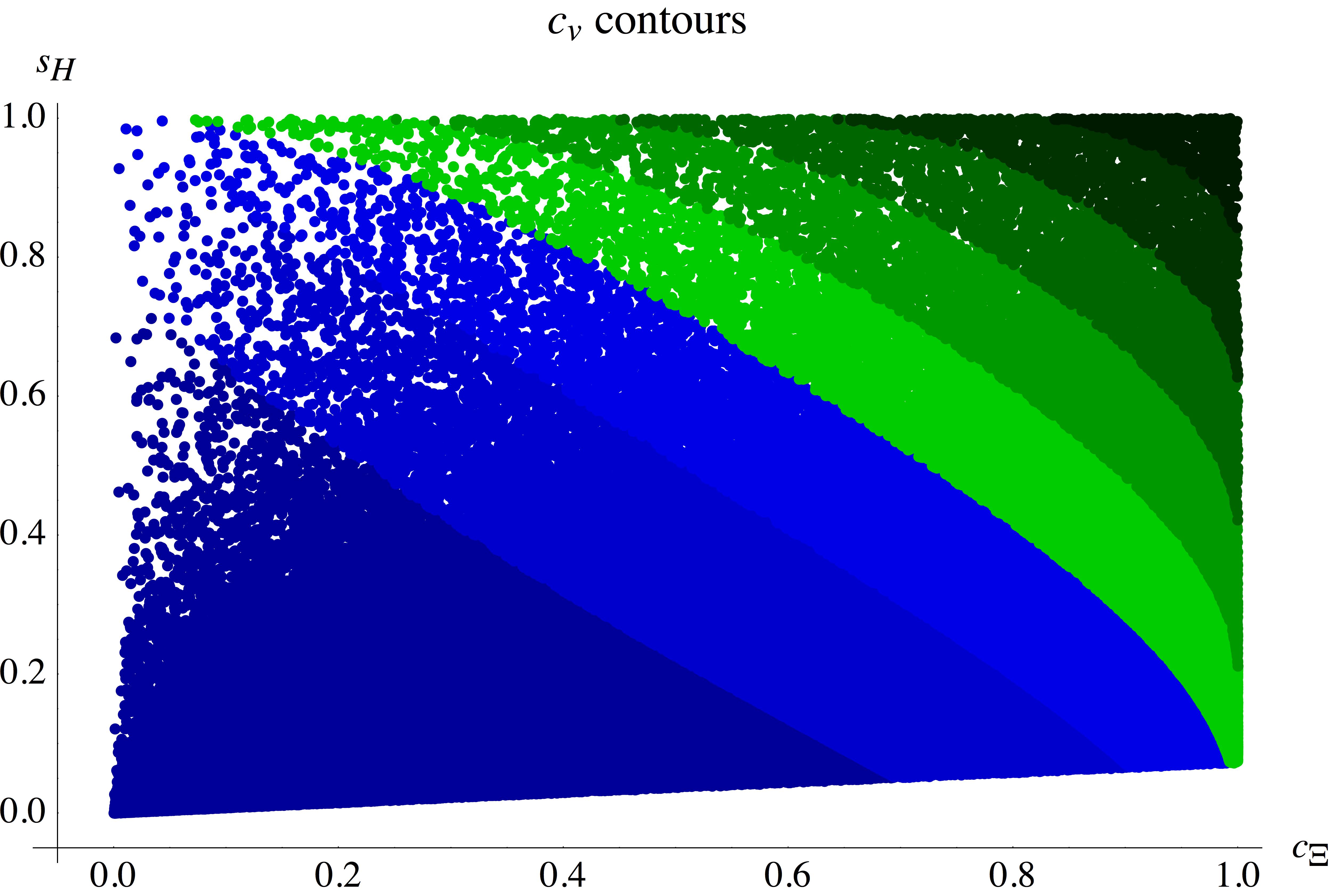}~
  \hspace{1.5cm}~
  \includegraphics[width=0.4\textwidth]{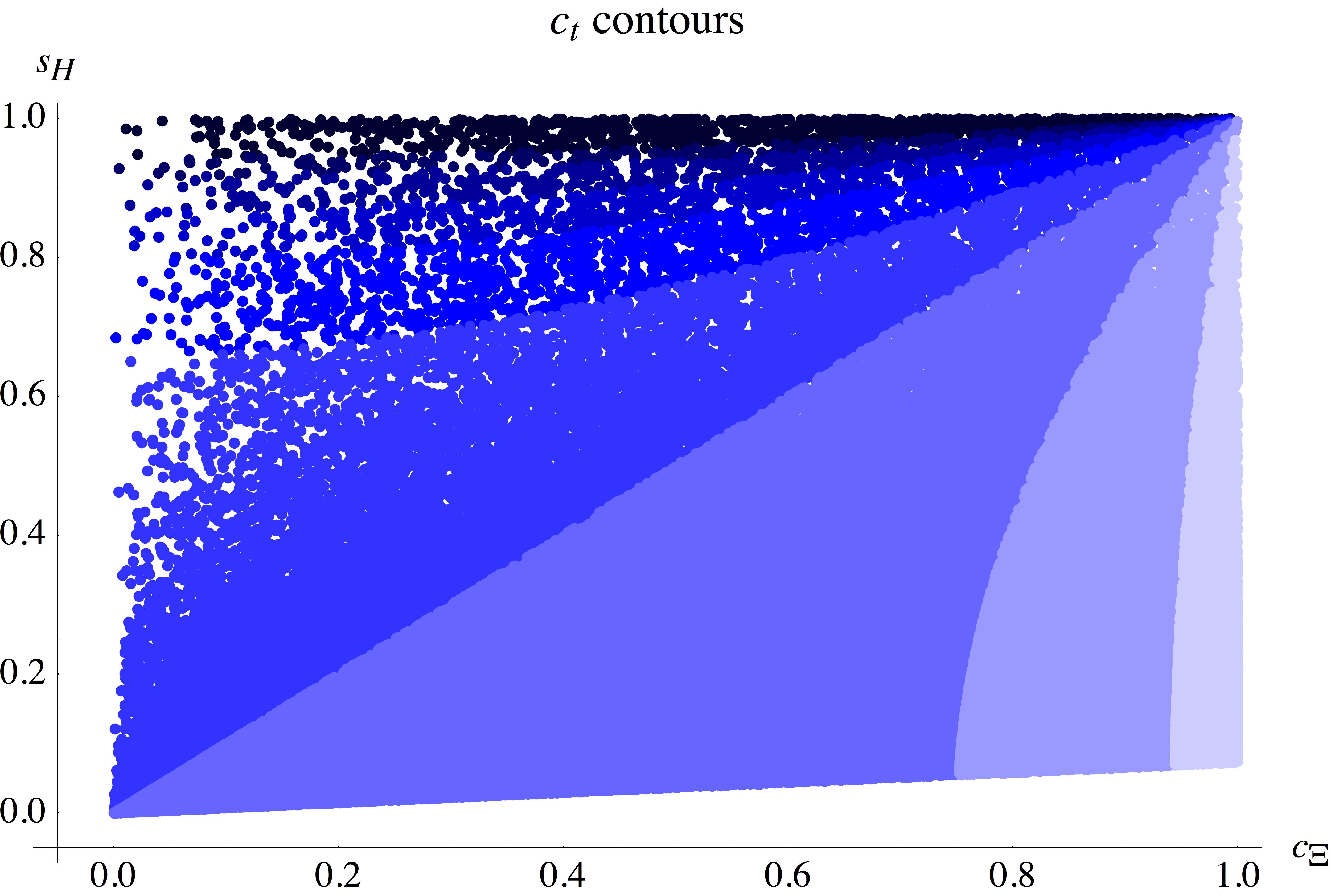}~
  \caption{ \label{fig:h0nosnot} Scan over the GM model's parameter
    space only requiring $m_{H_0}\simeq 126$~GeV in the
    $(c_\Xi,s_H)$-plane. The left panel shows $c_{v,H_0}$ contours with
    the following color codes: dark blue $-1<c_v<-0.66$, dark green
    $1.33<c_v< \sqrt{8/3}$. The right panel shows $c_{t,H_0}$ contours
    with dark blue $c_t<-3$, light blue $-0.33<c_t<0$.}
\end{figure*}
\begin{figure*}[p!]
  \includegraphics[width=0.5\textwidth]{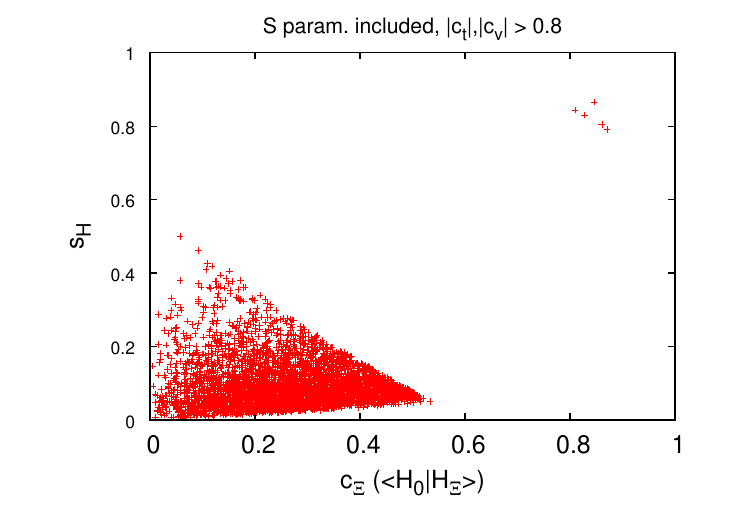}~
  \includegraphics[width=0.5\textwidth]{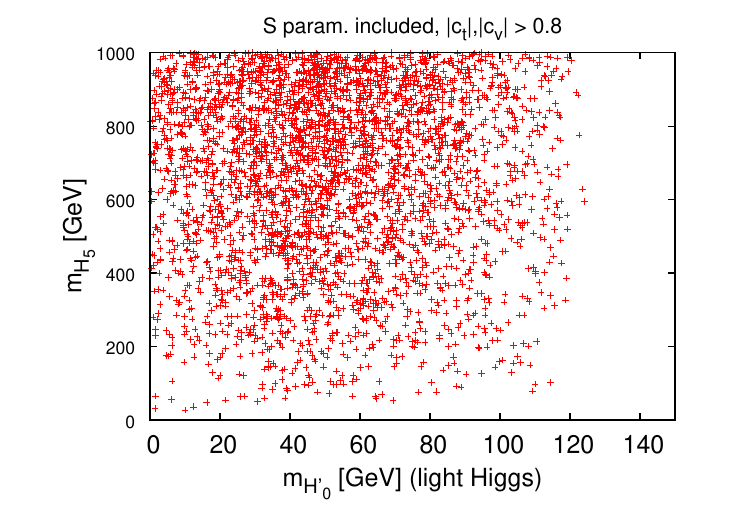}~
  \caption{ \label{fig:h0not} Scan of Fig.~\ref{fig:h0nosnot},
    including precision constraints on the $S$ parameter.}
\end{figure*}
\begin{figure*}[p!]
  \includegraphics[width=0.5\textwidth]{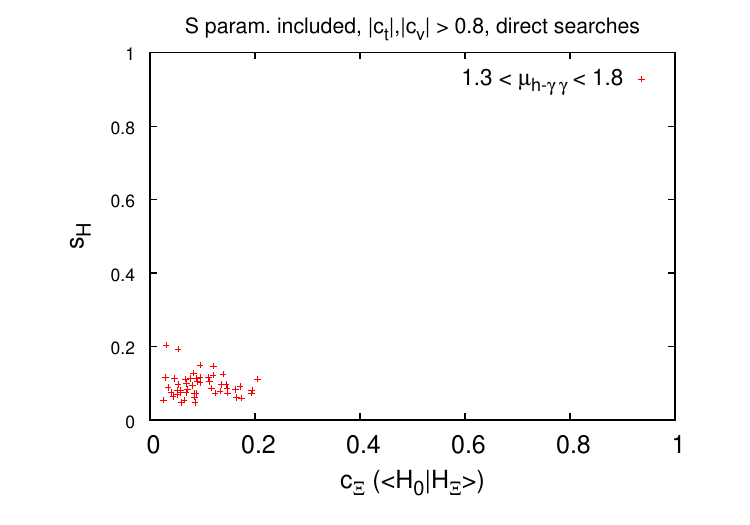}~
  \includegraphics[width=0.5\textwidth]{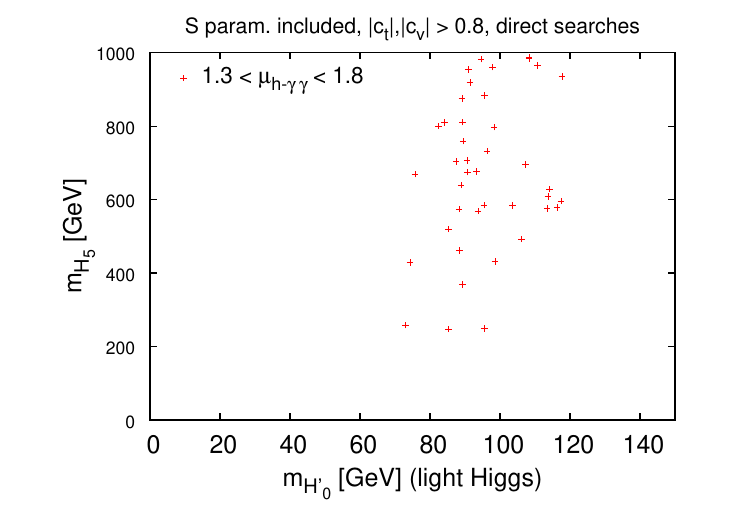}
  \caption{ \label{fig:h0notsearch} Scan of Fig.~\ref{fig:h0nosnot},
    including precision constraints on the $S$ parameter and signal
    strength constraints from direct searches.}
\end{figure*}

For later convenience, we also introduce the quantity 
\begin{equation}
  c_\Xi= \left\{ 
    \begin{matrix}
     \left\langle H'_0|H_\Xi\right\rangle = c_q, & \text{if}~m_{H'_0}\simeq 126~\text{GeV}\\
     \left\langle H_0|H_\Xi\right\rangle= s_q, & \text{if}~m_{H_0}\simeq 126~\text{GeV} 
    \end{matrix}
  \right.
\end{equation}
to quantify the overlap between the $126$~GeV mass eigenstate and
$H_\Xi$, the singlet whose mass would only be generated by $v_\Xi$ if
there was no mixing. 

At this point we also note that with the conventions used in this work
$c_t$ is always negative, whereas $c_v$ does not have a constrained
sign. Although being different to the conventions often used by other
groups, this is a perfectly legitimate choice that covers the
physically relevant cases. As a consequence, the SM-like situation
where the $hVV$ and $hf\bar{f}$ couplings have the same sign is
recovered in this work when $(c_t,c_v)=(-1,-1)$.

\subsection{${\bf{m_{H_0}\simeq 126 }}$~GeV --- inverted mass
  hierarchy}
\label{sec:noprime}
We first investigate the scenario where the Higgs candidate is more
closely related to the SM Higgs doublet, {\it i.e.} $m_{H_0}\simeq
126$~GeV.  This will also provide us the munition for the
phenomenologically more appealing case $m_{H'_0}\simeq 126$~GeV, for
which there are no constraints from Higgs decays to light states with
coupling strengths of the order of the weak scale.

\begin{figure}[!t]
\includegraphics[width=0.5\textwidth]{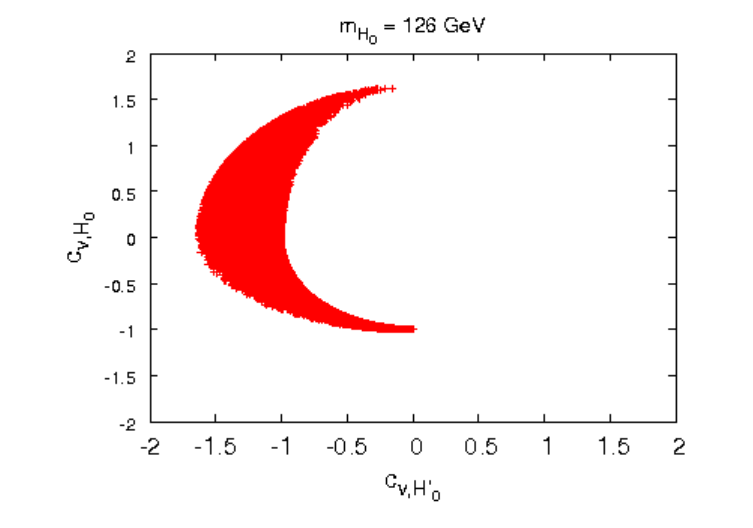}
\caption{\label{fig:cvhn} $(c_{v,H'_0}, c_{v,H_0})$ for $m_{H_0}\simeq
  126$~GeV. We impose neither electroweak precision constraints nor
  bounds from the signal strengths.}
\end{figure}

The scenario with $m_{H_0}\simeq 126$ GeV corresponds to a situation
which should be naively similar to the SM, because we expect that on
average $m_{H_0}\simeq 126$ GeV can be easily obtained when the triplet
vev is small, {\it i.e.} when Electroweak symmetry is mainly broken by the
doublet. However, as we mentioned above, in this scenario we have
$m_{H'_0}$ smaller than $m_{H_0}$, which obviously implies peculiar
consequences on the allowed phenomenology.

Fig.~\ref{fig:h0cvct} shows the model's couplings spans in the
$m_{H_0}\simeq 126$~GeV scenario, and in particular the allowed
enlarged range for $c_t$ and $c_v$ described by Eq.~\gl{eq:ctcv}.  We
start our walk through the constraints for $m_{H_0}\simeq 126$~GeV in
Fig.~\ref{fig:h0nosnot}, where no constraints have been imposed and
the isocontours for $|c_{v,H_0}|$ and $|c_{t,H_0}|$ are shown in the
$(c_\Xi,s_H)$ plane. In Fig.~\ref{fig:h0not} we impose the $S$
parameter constraint and in the panels of Fig.~\ref{fig:h0notsearch}
we further include the signal strengths of ATLAS. These steps sketch
the transition from LEP to the combined Higgs discovery and exclusion.

As expected, due to $m_{H'_0}\leq m_{H_0}$, we find large deviations
from the SM Higgs decay phenomenology, which is reproduced within
$\simeq 2\sigma$ by the current data. In particular, we find that this
feature holds even when we relax the constraints on the $T$ parameter:
for negative $c_v$ values, $|c_v|$ is always smaller than 1, in
particular when $|c_t|$ is different from 1. This can be understood
from the comments after Eq.~\eqref{eq:ctcv}: indeed, in this scenario,
$s_H$ is usually relatively small, whereas $c_\Xi\sim 0$, {\it i.e.}
$|c_q|\sim 1$. This means that there is room for $c_t$ to vary around
the central value (remaining however close to -1), whereas the vector
coupling essentially follows from $c_q c_H$. Therefore, $|c_v|$ is
bound to be smaller than 1. In particular, the more $c_t$ deviates to
-1, the smaller $|c_v|$ becomes. In such a situation, it is almost
impossible to reproduce the Higgs signal: we typically find values
$\mu_h\simeq 0.8$, whereas the value preferred by the excess observed
by ATLAS is $\simeq 1.25$. The few surviving points shown in
Fig.~\ref{fig:h0notsearch} correspond to $\mu_h\simeq 1$, which is
just within the $1\sigma$ error bar of the ATLAS global $\mu$
value. We also notice from the right plot in
Fig.~\ref{fig:h0notsearch} that for all these points the decay $H_0\to
H'_0H'_0$ is closed. On the other hand this means that a possible
future decrease in the observed $h\to \gamma \gamma$ rate can in
principle be accommodated by the GM triplet model in this scenario, at
the price of some tension with electroweak precision measurements.

Since the only constraint we have required is the $S$ parameter along
the lines of the previous section, it is natural to think that by
relaxing the $S$ parameter condition to $99\%$ confidence level, we
can find points in the parameter space with values of $c_t$ and $c_v$
that allow to reproduce the Higgs signal. Indeed, we observe that
there are points with larger values of $c_t$ and $c_v$ that survive
the relaxed electroweak precision constraints. They typically imply
larger values of $s_H$ and $c_\Xi$ values not necessarily close to 0.
Direct search constraints, however, both from LHC and LEP, remove
these points, typically because they violate the LEP bound on the
$HZZ$ coupling for the light Higgs state, as can be readily seen from
the left plot of Fig.~\ref{fig:h0cvct}: values of $c_{v,H_0}\ge 1$
correspond to values of $|c_{v,H'_0}|$ that are too large to survive
LEP bounds.

This scenario, seems to be heavily constrained by the $S,T,U$
parameters on top of the phenomenological requirements for the
observed signal strengths. The latter is expected from our previous
remarks on dominant decays to the lighter Higgs states, however, one
might naively expect that electroweak precision should not be too
constraining as this is definitely not the case for the model's limit
of the SM Higgs doublet with $m_h\simeq 126$~GeV \cite{lepepw}. So
what is the reason for electroweak precision observables being so
different from the SM in this case?

Let us step back and investigate how electroweak precision observables
are qualitatively influenced in the GM model. In comparison to the SM
the electroweak precision observables are influenced by the modified
Higgs sector. The gauge interactions of the fermions are unchanged and
by comparing to the SM reference point to calculate $\Delta S,\Delta
T, \Delta U$ drop out. To understand the Higgs-gauge boson
interactions which drive these observables via Eq.~\gl{eq:twopoint},
it is quite instrumental to understand how unitarity conservation is
realized in longitudinal gauge boson scattering in the GM
model. Looking at, {\it e.g.}, $W_LW_L$ scattering, a necessary
condition for unitarity to be conserved in any perturbative model is
that the coherent sum of new physics contributions to $WW\to WW$ has
to reproduce the SM Higgs contribution for high enough energies (above
all contributing thresholds). The $s$ and $t$-channel SM Higgs
exchange cancels the residual amplitudes growth proportional to the
the $W$'s energy squared in a minimal fashion. In the GM model this is
realized more intricately as, {\it e.g.}, in models with just simple
Higgs mixing. While in the latter case, for high enough energies, the
SM Higgs contribution is reproduced via $\sin^2\alpha +\cos^2\alpha=1$
($\alpha$ is the mixing angle), in the GM model we can have a very
large enhancement of $c_v$ by all uncharged Higgs particles to begin
with. Their ``over-contribution'' is cancelled by $t$- and $u$-channel
exchange of the $H^{\pm\pm}_5$ which in the high energy limit becomes
equivalent to an $s$-channel contribution~\cite{Gunion:1989ci}. Note
that all Higgs exchange diagrams are proportional to the squared real
Higgs couplings and the compensation results from kinematics $t,u\sim
-s$. The two-point functions of Eq.~\gl{eq:twopoint} obviously do not
encode any kinematics and are a coherent sum of the quartic Higgs
couplings and the trilinear couplings squared. Spontaneously broken
gauge invariance, on the other hand, guarantees the absence of UV
singularities via cancellations among the different contributing
diagrams (and the corresponding $T$ parameter counter term).

\begin{figure}[b!]
  \includegraphics[width=0.5\textwidth]{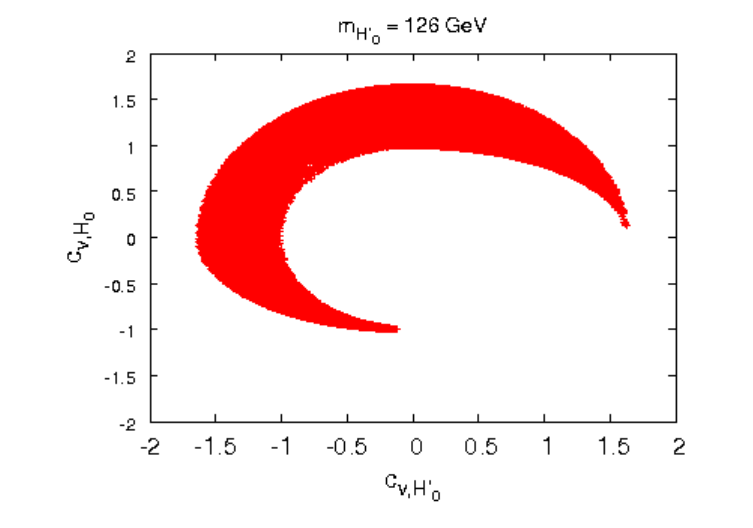}
  \caption{\label{fig:cvhpn} $(c_{v,H'_0}, c_{v,H_0})$ for
    $m_{H'_0}\simeq 126$~GeV. We impose neither electroweak precision
    constraints nor bounds from the signal strengths.}
\end{figure}

From Fig.~\ref{fig:cvhn} we can therefore immediately read off the
potential issue of the GM model that arises when it is confronted with
electroweak precision measurements for $m_{H_0}\simeq 126$~GeV.  There
we plot the $(c_{v,H'_0}, c_{v,H_0})$ correlation for our scan with
the requirement of $H_0$ being the observed Higgs-like
candidate. Obviously there is no anti-correlation of the two Higgs
states. When enforcing the observed data's constraint on
$|c_{v,H_0}|\simeq 1$ (horizontal lines in Fig.~\ref{fig:cvhn}) we
typically have sizeable values for $c_{v,H'_0}$: therefore $\Delta S$
generically turns out to be large when we require the Higgs
candidate's couplings to reproduce the SM or to obtain even larger
couplings than the SM, $c_{v,H_0}\gtrsim 1$. This together with the
large deviation of the Higgs phenomenology driven by
${\text{BR}}(H_0\to H'_0 H'_0)$ highly constrains $m_{H_0}\simeq
126$~GeV, independent of a possible excess in ${\text{BR}}(H_0\to
\gamma\gamma)$.

\begin{figure*}[!t]
  \includegraphics[width=0.45\textwidth]{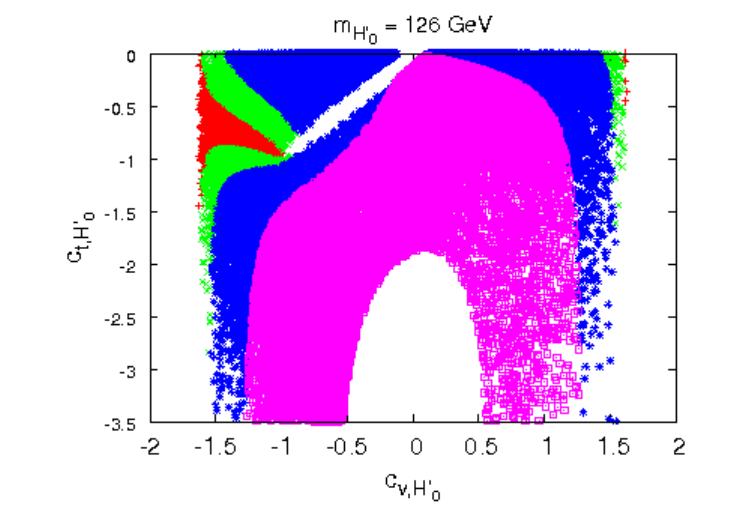}
  \includegraphics[width=0.45\textwidth]{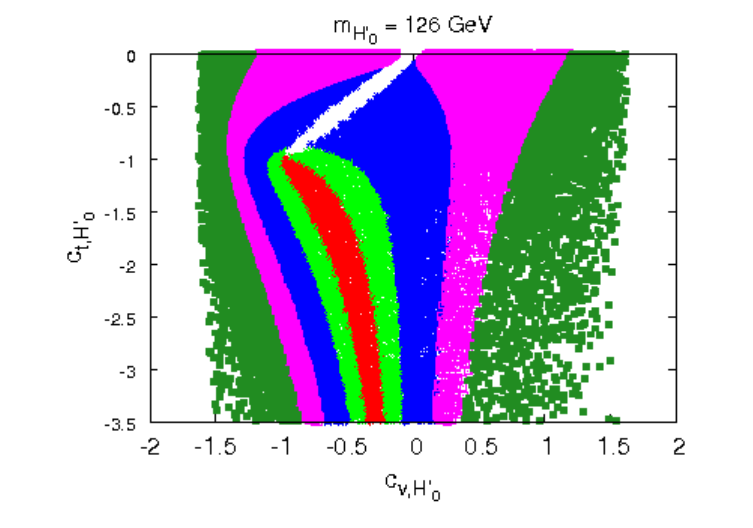}
  \caption{\label{fig:h0pcvct} $(c_{v,H'_0},c_{t,H'_0})$ correlation
    for $m_{H'_0}\simeq 126$~GeV. We do not impose any other additional
    constraints. The color code is the same as in
    Fig.~\ref{fig:h0cvct}, although here contours are for $c_{v,H_0}$
    (on the left panel) and $c_{t,H_0}$.}
\end{figure*}

\subsection{${\bf{m_{H'_0}\simeq 126 }}$~GeV --- normal mass
  hierarchy}
\label{sec:prim126}

We turn to identifying $H'_0$ with the observed Higgs candidate. In
Fig.~\ref{fig:h0pcvct} we plot the different contour regions for
$c_{v,H_0}$ and $c_{t,H_0}$ as functions of $(c_{v,H'_0},c_{t,H'_0})$
without imposing any constraints apart from $m_{H'_0}\simeq
126$~GeV. The left-out region inside the blue contours stems from
parameter points that do not give rise to (local) minimum of the
potential. The model clearly predicts large enhancements of the
vectorial couplings, which also manifest in an enhanced partial decay
width $\Gamma(H'_0\to \gamma \gamma)$, which is a function of basic
contributions: the $W$ loop, the fermion loops and the charged scalar
loops. In the SM the $W$ loop dominates over the top
contributions. Fig.~\ref{fig:h0pcvct} indicates that this correlation
is already significantly altered via Eq.~\gl{eq:ctcv}: the top loop
can be suppressed while the $W$ loop is enhanced. This needs to be
contrasted to ordinary complex triplet models, where, due to the small
triplet vev, such an enhancement via $c_v$ is not present and possible
branching ratio enhancements need to be driven essentially only by the
additional charged scalars.

\begin{figure*}[p!]
  \includegraphics[width=0.4\textwidth]{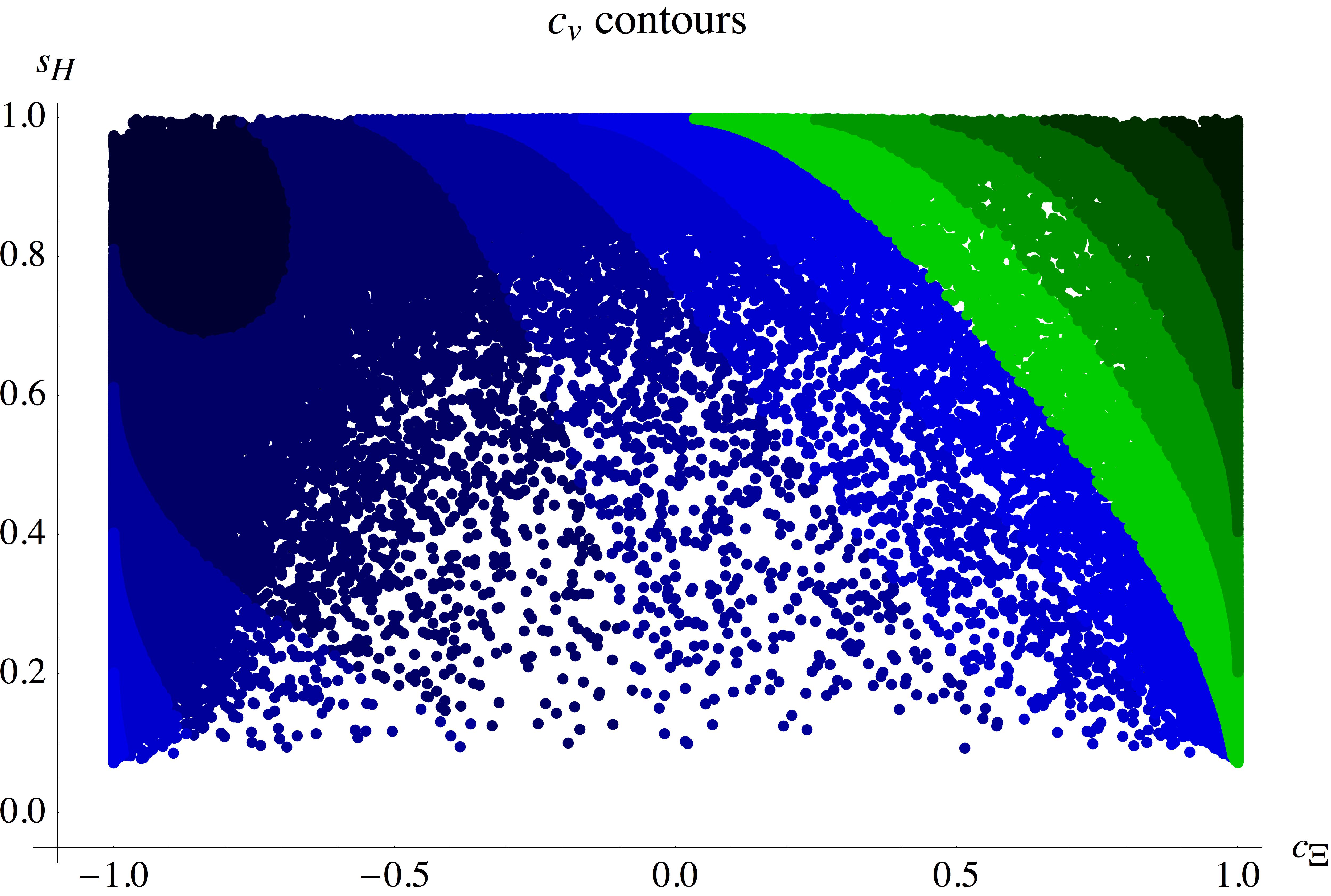}~
  \hspace{1.5cm}~
  \includegraphics[width=0.4\textwidth]{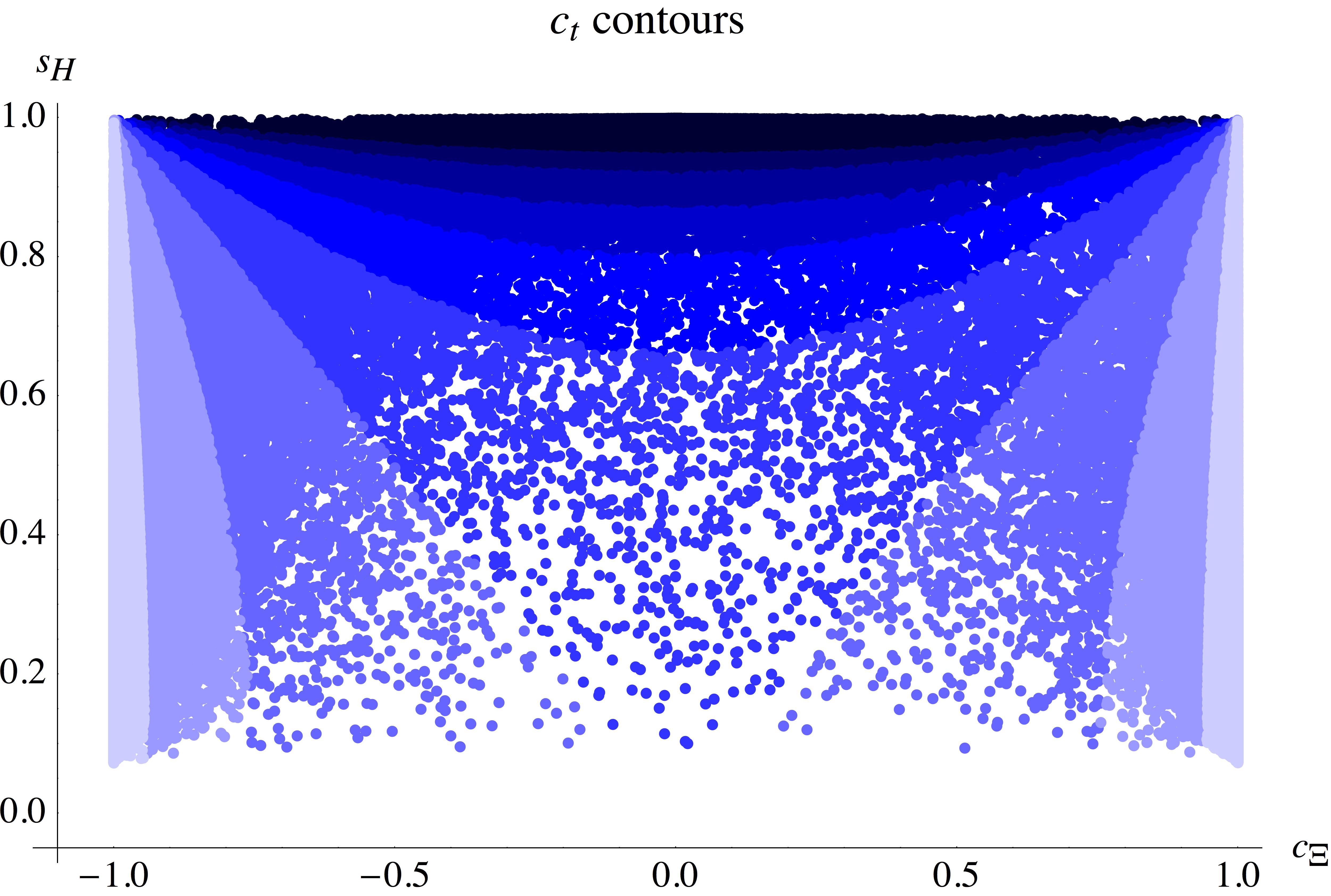}~
  \caption{ \label{fig:h0pnosnot} Scan over the GM model’s parameter
    space only requiring $m_{H'_0}\simeq 126$ GeV in the
    $(c_\Xi,s_H)$-plane. The left panel shows $c_{v,H'_0}$ contours
    with the following color codes: dark blue $-\sqrt{8/3}<c_v<-1.33$,
    dark green $1.33<c_v< \sqrt{8/3}$. The right panel gives
    $c_{t,H'_0}$ contours with dark blue $c_t<-3$, light blue
    $-0.33<c_t<0$.}
\end{figure*}
\begin{figure*}[p!]
  \includegraphics[width=0.5\textwidth]{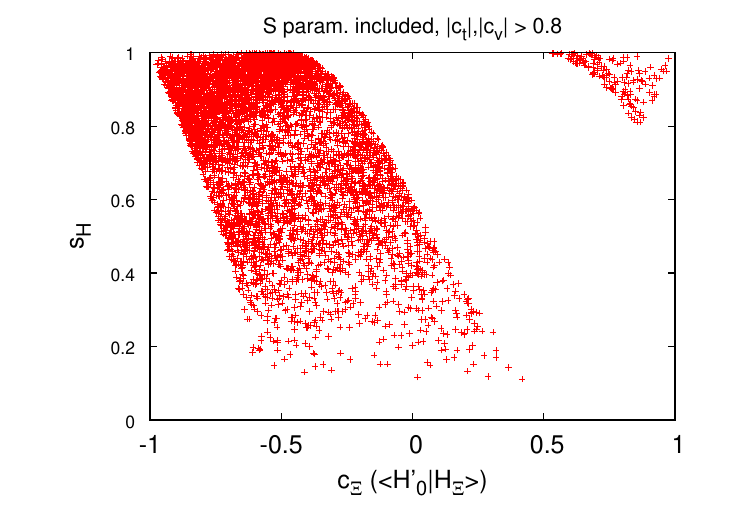}~
  \includegraphics[width=0.5\textwidth]{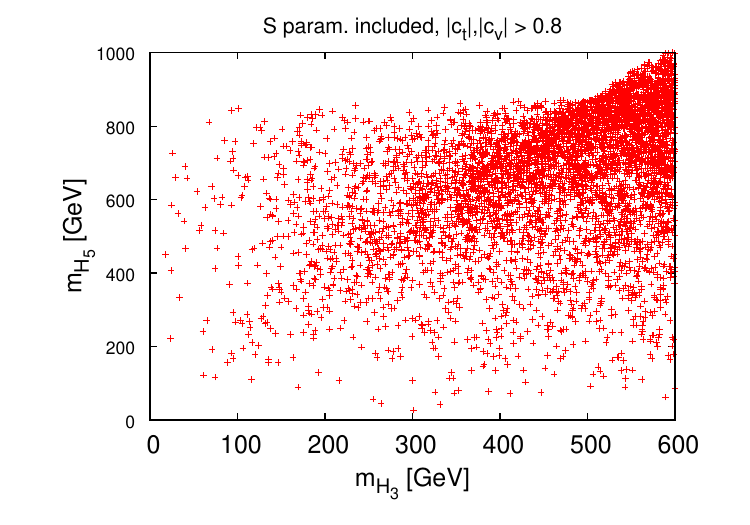}~
  \caption{ \label{fig:h0pnnot} Scan of Fig.~\ref{fig:h0pnosnot},
    including precision constraints on the $S$ parameter.}
\end{figure*}
\begin{figure*}[p!]
  \includegraphics[width=0.5\textwidth]{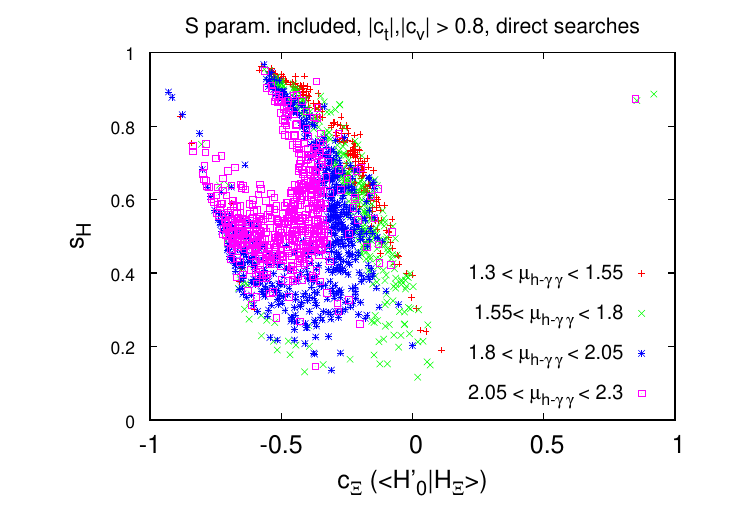}~
  \includegraphics[width=0.5\textwidth]{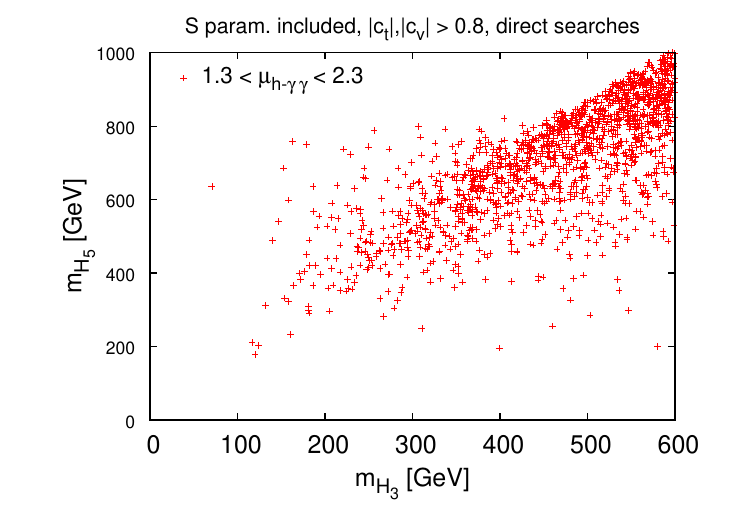}~
  \caption{ \label{fig:h0pnnotsearch} Scan of
    Fig.~\ref{fig:h0pnosnot}, including precision constraints on the
    $S$ parameter and signal strength constraints from direct
    searches.}
\end{figure*}

\begin{figure*}[p!]
  \includegraphics[width=0.5\textwidth]{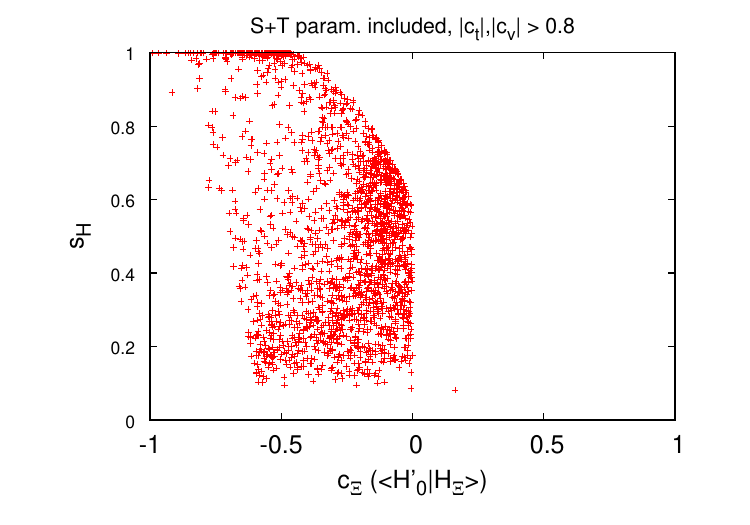}~
  \includegraphics[width=0.5\textwidth]{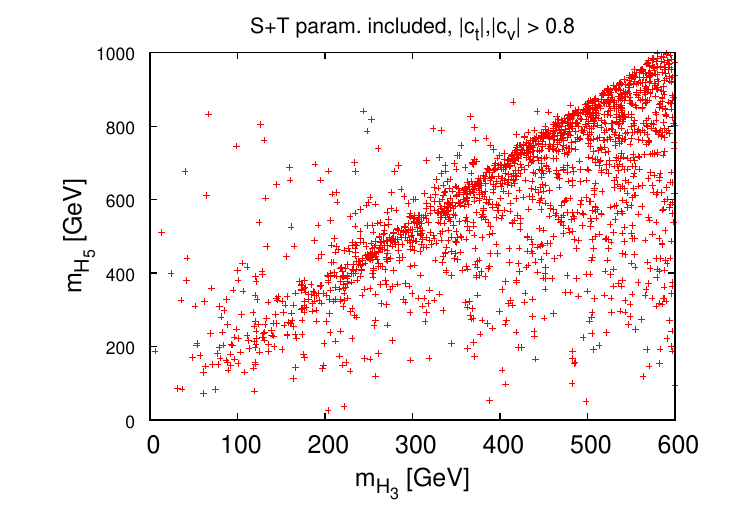}~
  \caption{ \label{fig:h0pnt} Scan of Fig.~\ref{fig:h0pnosnot},
    including precision constraints on both the $S$ and the $T$
    parameters.}
\end{figure*}
\begin{figure*}[p!]
  \includegraphics[width=0.5\textwidth]{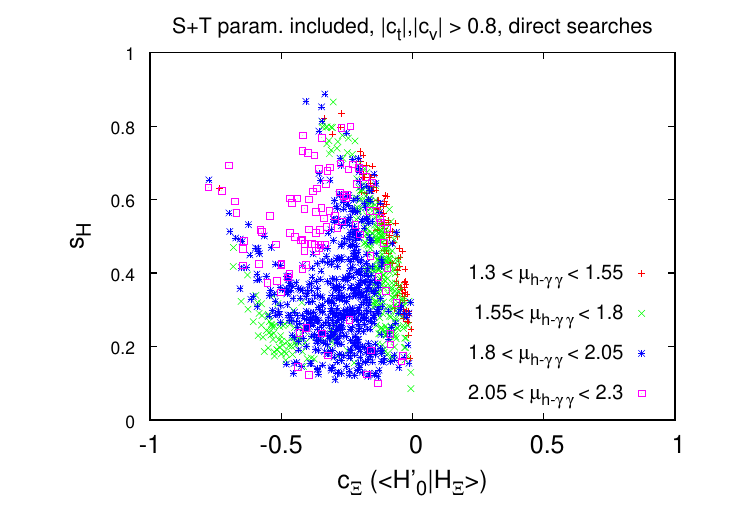}~
  \includegraphics[width=0.5\textwidth]{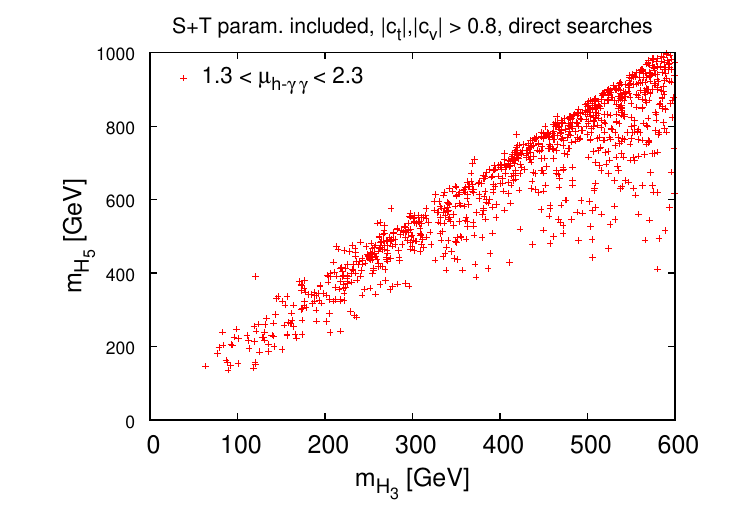}~
  \caption{ \label{fig:h0pntsearch} Results of
    Fig.~\ref{fig:h0pnnotsearch} including constraints on the $S$ and
    $T$ parameters, and signal strength constraints from direct
    searches.}
\end{figure*}
\begin{figure*}[p!]
  \includegraphics[width=0.5\textwidth]{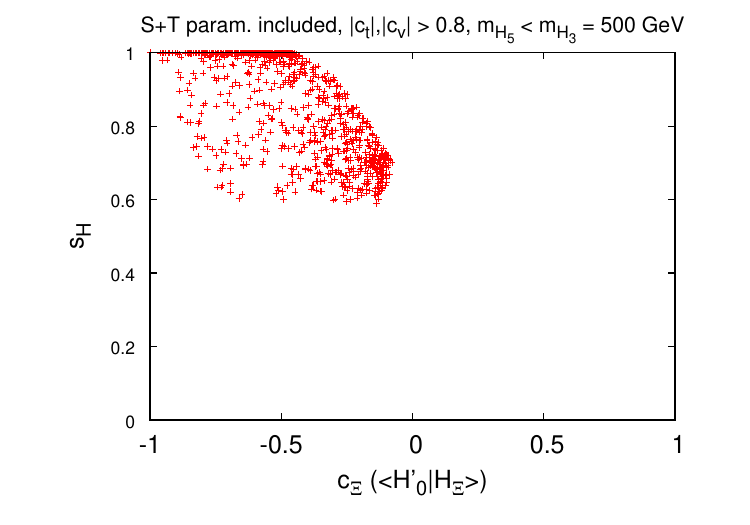}~
  \includegraphics[width=0.5\textwidth]{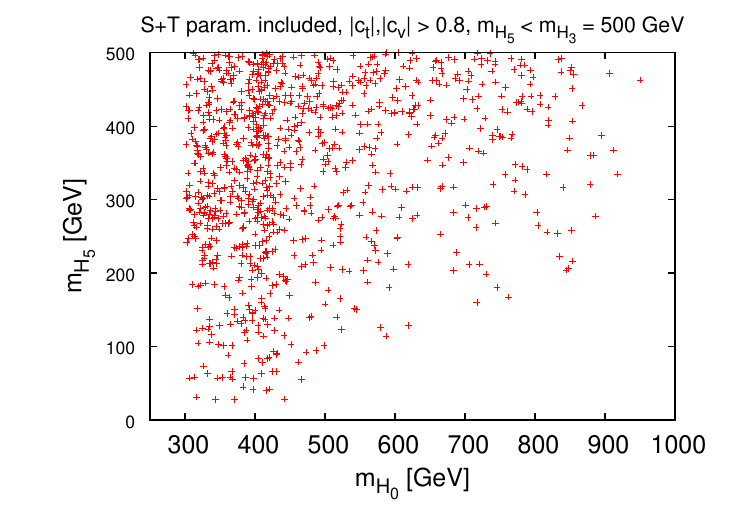}~
  \caption{ \label{fig:h0pn35t} Dedicated scan with $m_{H_3}=500$~GeV
    and $m_{H_5}< m_{H_3}$ while $m_{H'_0}\simeq 126$~GeV. Constraints
    on both the $S$ and $T$ parameters are included.}
\end{figure*}

The contribution from {\it e.g.} charged $H_5^\pm$ and $H_5^{\pm\pm}$
scalars in the GM model to the $H'_0 \to \gamma \gamma$ amplitude
scales as $F(s_H,\vec{\lambda})Q_{h_5}^2 [v_{\text{SM}}^2/(2m_5^2)]
A_0(m_{H'_0}^2/4m_{H_5}^2)$, where $F$ is a function of the parameter
space and $A_0$ is the typical one-loop 3-point function encountered
in such amplitudes~\cite{gf}. A similar relation holds for
$H^\pm_3$. The 3-point function is essentially constant for masses
larger than $m_{H_5}\gtrsim 200$ GeV, $A_0\simeq 0.35$. In the end the
charged scalar effect can be compensated by the vast parameter range
that is admissible due to changes in $(c_v,c_t)$, which can also
constructively interfere as a consequence of Eq.~\gl{eq:ctcv}. We find
that the $H_5$ contribution is typically small and destructive over
the parameter range that we consider, {\it i.e.} it works against the
$W$ contribution; this means that a necessary condition to have
$\text{BR}(H'_0\to \gamma\gamma)\gtrsim \text{BR}(H'_0\to
\gamma\gamma)_{\text{SM}}$ is that the destructive interference of the
SM $t$ and $W^\pm$ is reduced or becomes constructive.

We again start our walk through the constraints for $m_{H'_0}\simeq
126$~GeV in Fig.~\ref{fig:h0pnosnot}. In Fig.~\ref{fig:h0pnnot} we
impose the $S$ parameter constraint and in the panels of
Fig.~\ref{fig:h0pnnotsearch} we include the signal strengths of
ATLAS. We show the fully tuned setup in the analogous
Figs.~\ref{fig:h0pnt} and \ref{fig:h0pntsearch}. From
Fig.~\ref{fig:cvhpn} we expect the $m_{H'_0}\simeq 126$~GeV scenario
to be less sensitive to electroweak precision constraints:
$|c_{v,H'_0}|\sim 1$ (vertical lines in Fig.~\ref{fig:cvhpn}) allows
for a wide range on $c_{v,H_0}$.

\begin{figure*}[t!]
  \includegraphics[width=0.5\textwidth]{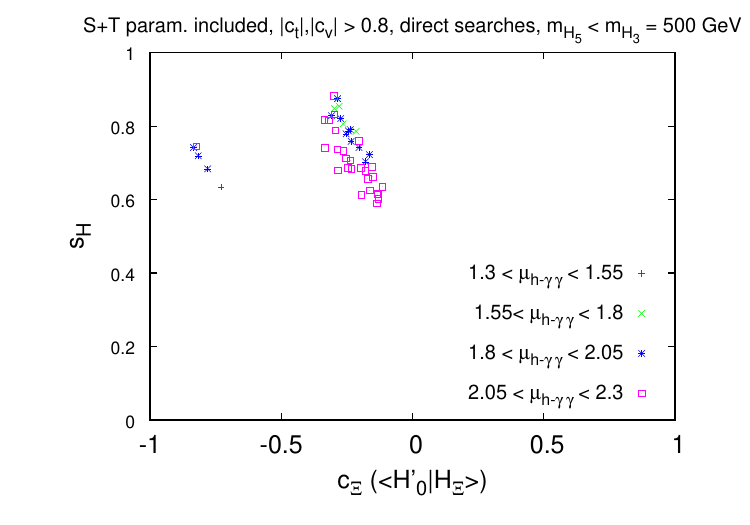}~
  \includegraphics[width=0.5\textwidth]{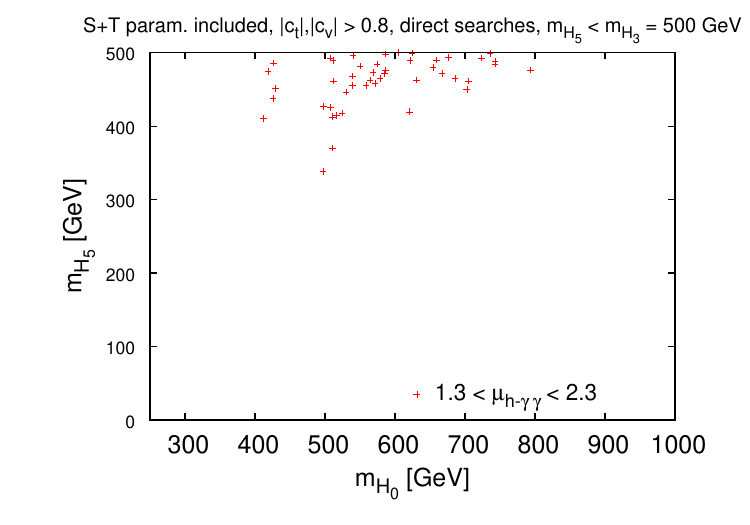}~
  \caption{ \label{fig:h0pn35tsearch} Signal strengths included on top
  of the results of Fig.~\ref{fig:h0pn35t}.}
\end{figure*}

Indeed, by enforcing electroweak precision constraints we find our
results to be also less sensitive to the $T$ parameter, although
more parameter points at large $s_H$ are rejected. Nonetheless we
still obtain consistent parameter choice for large values of $s_H$,
which as consequence of signal strengths then needs to be
anti-correlated to $c_\Xi$. The projection on $(c_{\Xi},s_H)$ therefore
nicely discriminates between different regions of diphoton branching
ratio enhancements. We again see that a potential excess in the
diphoton final state in the GM model is predominantly realized via
relaxing the SM-constraints on couplings to the top and massive gauge
boson sector with the additional scalars providing some additional
freedom in this respect, yet their impact is subdominant unless
$m_{H_5}$ is significantly small.

A generic prediction of the GM model when it reproduces the current
data and does not violate electroweak constraints on $S,T$ is the
presence of a relatively heavy quintet. The triplet state are also
heavy but their phenomenology is sufficiently suppressed (keep in mind
that we impose ATLAS exclusion contours on the uncharged states
throughout). The quintet's production is suppressed but given the mass
of this state we expect a measurement of the phenomenologically clean
same-sign $W$ production via WBF to give substantial constraints on
the realization $m_{H'_0}\simeq 126$~GeV. Adapted searches will
therefore have the potential to further constrain the GM model's
allowed parameter space, or, conversely, to find an hint of a doubly
charged scalar.

\subsubsection*{Heavy triplet --- light quintet}
Until now our parameter choices are dominated by choices such that the
quintet states typically outweigh the triplet Higgs bosons. Now, we
specifically analyze the situation when the triplet is heavy
($m_{H,3}\simeq 500$~GeV) and the quintet states are light
$m_{H_5}<m_{H_3}$. The resulting spectra with a heavy triplet and a
relatively light quintet should be favored by flavor analyses, since
the quintet is fermiophobic and the triplet is less related to fermion
mass generation and should hence decouple for large masses from flavor
observables, when $s_H$ is not too large. We will also get a better
understanding of the correlation between $s_H$ and $c_q$.

As shown in Figs.~\ref{fig:h0pn35t} and \ref{fig:h0pn35tsearch}, this
dedicated scan projects out the values of $s_H\simeq 1$ with typically
heavy additional singlets $H_0$ in the range of several 100
GeV. With respect to our previous remarks in Sec.~\ref{sec:prim126}
about how $H'_0\to \gamma\gamma$ comes about, we do not find any
notable qualitative modification of our earlier findings: the enhanced
diphoton branching is still predominantly enabled via the interplay of
$c_{v,H'_0}$ and $c_{t,H'_0}$ and potentially large branching ratios
can be achieved. These parameter choices typically imply a very large
$H_0$ mass compared to the other case -- future searches for uncharged
SM Higgs-like states in the hundreds of GeV range will highly
constrain this fact of the GM model.

\subsubsection*{SM-like GM phenomenology}
\begin{figure*}[t!]
  \includegraphics[width=0.5\textwidth]{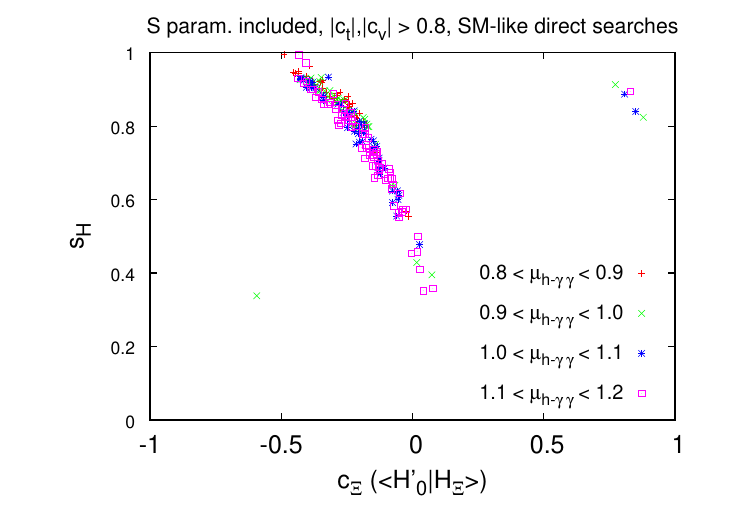}~
  \includegraphics[width=0.5\textwidth]{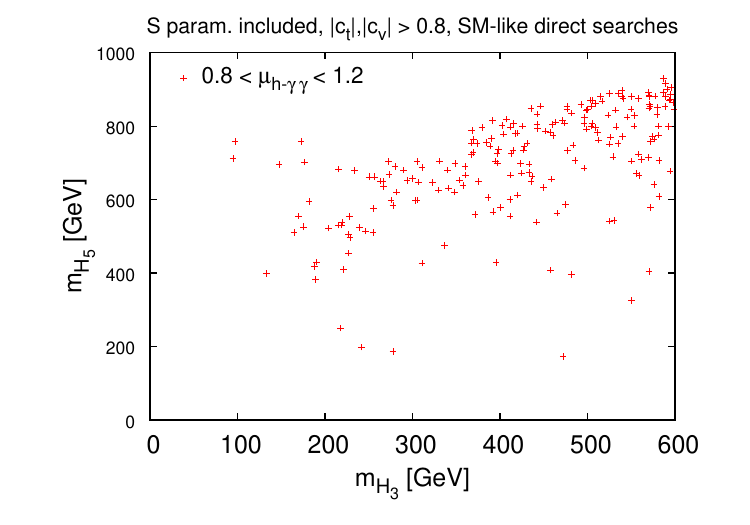}~
  \caption{ \label{fig:h0pnnotsearchSM} Scan of
    Fig.~\ref{fig:h0pnosnot}, including precision constraints on the
    $S$ parameter, and signal strength constraints, in a SM-like
    scenario where $\mu_{h\to WW}$, $\mu_{h\to\gamma\gamma}$ and the
    total $\mu$ value lie in the interval [0.8,1.2].  }
\end{figure*}

\begin{figure*}[t!]
  \includegraphics[width=0.5\textwidth]{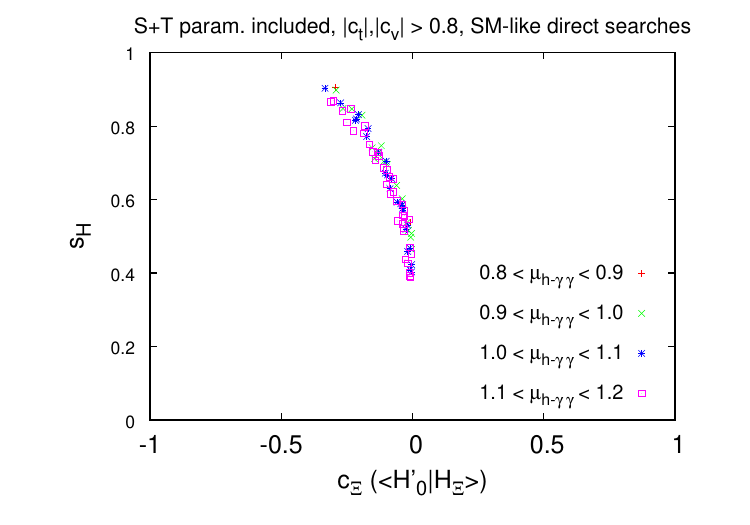}~
  \includegraphics[width=0.5\textwidth]{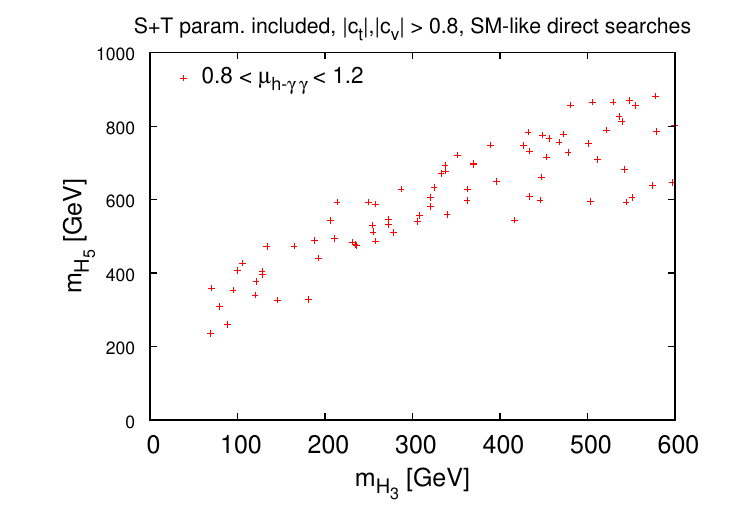}~
  \caption{ \label{fig:h0pntsearchSM} Scan of
    Fig.~\ref{fig:h0pnosnot}, including precision constraints on the
    $S$ and $T$ parameters, and signal strength constraints, in a
    SM-like scenario where $\mu_{h\to WW}$, $\mu_{h\to\gamma\gamma}$
    and the total $\mu$ value lie in the interval [0.8,1.2].  }
\end{figure*}

Let us abandon the diphoton excess initially observed by both ATLAS
and CMS and briefly investigate the parameter range for which
${\text{BR}}(H'_0\to \gamma\gamma) \simeq
{\text{BR}}(h\to\gamma\gamma)_{\text{SM}}$. This amounts to a
potential future measurement which shows an even larger resemblance to
the SM than we currently observe. Parameter choices of the GM model
with such a measurement would be required to be more tailored to the
SM. Identifying $H_0$ as our Higgs candidate is obviously still
disfavored along the lines of our discussion of
Sec.~\ref{sec:noprime}. Furthermore, from our analysis of the
$m_{H'_0}\simeq 126$~GeV it should be clear that the number of degrees
of freedom in the light of not too constraining Higgs searches for
masses different than $126$~GeV and the available parameter space are
large enough to account for such a situation (a limit of $\sigma\times
\Br / [\sigma\times \Br]_{\text{SM}} \simeq 0.1$ roughly corresponds
to $|c_v|\lesssim 0.32$). This is especially true for parameter
choices for which an additional scalar falls into the vicinity of the
observed $126$~GeV Higgs candidate, where resolution effects limit
stringent constraints. We assume consistency with the SM hypothesis
within $20\%$ of the $WW,ZZ,\gamma\gamma$ and combined categories in
the following, but we keep the ATLAS constraints on the other Higgs
states unmodified. With our remarks about the $(c_\Xi,s_H)$ plane
being a good discriminant of the diphoton enhancement in mind we
observe in Fig.~\ref{fig:h0pnnotsearchSM} and~\ref{fig:h0pntsearchSM}
that the SM-like requirements slice out a specific parameter region in
the GM model (independent of the $T$ parameter).

This teaches two important lessons for an observation of perfectly
consistent SM Higgs measurements in the future. On the one hand, even
if such an outcome does not speak in favor of the GM model, there is
still a lot of parameter space available, which can and needs to be
tested at the LHC. From this point of view, it again appears
indispensable to extend existing Higgs-like searches to the heavy mass
regime while relaxing specific assumptions on the total particle width
as limiting factor of such analyses. With again quintet masses in the
hundreds of GeV regime, which can be straightforwardly accessed at the
LHC in same-sign $W$ final states, the GM model can be highly
constrained, even when its Higgs candidate phenomenology highly
resembles the SM Higgs.


\section{Conclusions}
\label{sec:conc}
The Georgi-Machacek model implements Higgs triplets in a custodially
invariant way at the price of an additional fine tuning problem. In
this paper we have performed an analysis of the model's parameter
space in the light of direct collider and electroweak precision
constraints. Depending on the interpretation of electroweak precision
in the GM model context, there are important consequences for the model
when it is confronted with data. Given that the GM model has a wide
range of coupling spans, excesses in the $h\to \gamma \gamma$ rate can
be observed, but there is also a vast parameter region that is allowed
if data resembles the SM.

In addition to the SM, the GM model introduces a number of new
scalars. However, in the light of recent measurements only a subset of
these can be considered the discovered Higgs candidate. One
peculiarity that arises in the GM model is that if the Higgs boson
that arises mostly from the SM-like doublet Higgs field is identified
with the Higgs candidate at around 126 GeV, there is a light scalar in
the spectrum which mostly arises from the triplet. This puts strong
constraints on this option as branching ratios to SM matter quickly
deviate from their observed values when the light scalar becomes
accessible as a Higgs decay channel. Furthermore, electroweak
precision constraints (mostly by the $S$ parameter) disfavor this
option even if the light scalar is not constrained by LEP
measurements. If the excess in the diphoton channel prevails this
option becomes heavily constrained whereas lower combined signal
strengths can be obtained at the cost of a tension with electroweak
precision data.

Most of these constraints are relaxed when we identify the non-doublet
state with the observed candidate. Here a vast parameter space becomes
available to accommodate excesses in the $\gamma\gamma$ channel, but
the model also exhibits parameter regions consistent with SM.

For our simplified lagrangian approach, the consistent parameter
choices of the SM typically predict the presence of new states in the
hundreds of GeV region. Lighter spectra are admissible by current
collider constraints but typically result from extremely small mixing
of the custodial triplets, one of which giving rise to massive gauge
bosons.  While the triplet mass eigenstate is \CP~odd and can only be
produced via gluon fusion, the quintet state is fermiophobic and can
be constrained via a dedicated measurement in WBF-type production in
the near future. The latter search is likely to put extremely tight
constraints on the GM model, especially if future measurements of the
126 GeV Higgs candidate are consistent with the SM predictions.

\acknowledgements
We would like to thank Claude
Duhr for {\sc{FeynRules}} support.  CE thanks Adam Falkowski and Jure
Zupan for discussions during the CERN BSM Theory Summer Institute. ER
thanks Uli Haisch for useful discussions. 

CE acknowledges funding by the Durham International Junior Research
Fellowship scheme.


\end{document}